\documentclass[aps,prd,eqsecnum,11pt,showpacs,preprintnumbers,nofootinbib]{revtex4}
\usepackage{graphicx}
\usepackage{subfig}
\usepackage{amssymb,amsmath}
\newcommand{\bes}{\begin{subequations}}
\newcommand{\ees}{\end{subequations}}
\newcommand{\be}{\begin{equation}}
\newcommand{\ee}{\end{equation}}
\newcommand{\bea}{\begin{eqnarray}}
\newcommand{\eea}{\end{eqnarray}}
\newcommand{\nn}{\nonumber}
\newcommand{\ern}{Extreme Reissner-Nordstr\"{o}m}
\def\frc#1#2{{\textstyle{{#1 \over #2}}}}
\begin{document}

\title{Semiclassical Gravity in the Far Field Limit of Stars, Black Holes, and Wormholes}
\author{Eric D. Carlson$^{\rm a}$}
\altaffiliation{\tt ecarlson@wfu.edu}
\author{Paul R. Anderson$^{\rm a}$}
\altaffiliation{\tt anderson@wfu.edu}
\author{Alessandro Fabbri$^{\rm b,c}$}
\altaffiliation{\tt afabbri@ific.uv.es}
\author{Serena Fagnocchi$^{\rm d}$}
\author{William H. Hirsch$^{\rm a}$}
\author{Sarah A. Klyap$^{\rm a}$}
\affiliation{${}^{a)}$Department of Physics, Wake Forest University,
Winston-Salem, North Carolina 27109, USA,}
\affiliation{$^{b)}$ Departamento de F\'{\i}sica Te\'orica and IFIC,
Universidad de Valencia-CSIC, Calle Dr.\ Moliner 50, Burjassot 46100, Valencia, Spain.}
\affiliation{$^{c)}$ APC (Astroparticules et Cosmologie), Universit\'e Paris 7 - Denis Diderot, 10 rue Alice Domon et L\'eonie Duquet, 75205 Paris Cedex 13, France}
\affiliation{$^{d)}$School of Physics and Astronomy, University of Nottingham, University Park, Nottingham NG7 2RD, UK.}
\begin{abstract}

Semiclassical gravity is investigated in a large class of asymptotically flat, static, spherically symmetric spacetimes
including those containing static stars, black holes, and wormholes.  Specifically the stress-energy tensors of massless free spin $0$ and spin $\frc12$
fields are computed to leading order in the asymptotic regions of these spacetimes.  This is done for spin $0$ fields in Schwarzschild spacetime using
a WKB approximation.  It is done numerically for the spin $\frc12$ field in Schwarzschild, extreme Reissner-Nordstr\"{o}m, and various wormhole spacetimes.
And it is done by finding analytic solutions to the leading order mode equations in a large class of asymptotically flat static spherically symmetric spacetimes.
Agreement is shown between these various computational methods.  It is found that for all of the spacetimes considered, the energy density and pressure in the asymptotic region are proportional
to $r^{-5}$ to leading order.  Furthermore, for the spin $1/2$ field and the conformally coupled scalar field, the stress-energy tensor depends only on the leading order geometry in the far field limit.  This is also true for the minimally coupled scalar field for spacetimes containing either a static star or a black hole, but not for spacetimes containing a wormhole.

\end{abstract}


\maketitle

\newpage

\section{Introduction}

A number of studies have been carried out of the leading order quantum corrections to the gravitational potential due to
quantized fields in static spherically symmetric spacetimes in the weak field limit~\cite{duff,donoghue,h-l,mazz,mazz2,woodard,more}.  In two cases~\cite{mazz,mazz2} the leading order behaviors of the energy densities and pressures of the quantized fields were also computed.
In this limit it is assumed that the gravitational field is everywhere weak such as in the
case of a planet or a nonrelativistic star.  The far field limit in an asymptotically flat spacetime is similar to the weak field limit in that, in the asymptotic region, the spacetime is nearly flat.  The difference is that the spacetime curvature need not be small everywhere.
Since quantum field theory is intrinsically nonlocal, it is in principle more difficult to
study quantum effects in the far field limit when the spacetime curvature is not small everywhere and the fields are in the vacuum state which is natural for the asymptotically flat region, often called the Boulware state~\cite{boulware}.\footnote{This statement does not apply to thermal states such as the Hartle-Hawking-Israel state~\cite{hh}, because for such states the leading order behavior of the stress-energy tensor for a quantized field in the far field region is the same as it would be in Minkowski space.}
In fact there are only two previous calculations that we are aware of which focus on the far field limit:
Anderson, Balbinot, and Fabbri~\cite{abf} computed the leading order asymptotic behavior for the stress-energy tensor, $\langle {T^a}_b \rangle$ for massless scalar fields with arbitrary coupling $\xi$ to the scalar curvature in the Boulware state in Schwarzschild spacetime; and  Garbarz, Giribet, and Mazzitelli~\cite{mazz2} calculated for the same fields in $D$ dimensional Schwarzschild-Tangherlini spacetime with $D\ge 4$ the leading order asymptotic behaviors of both $\langle \phi^2 \rangle$ and $\langle {T^a}_b \rangle$.

An interesting question that can be addressed by comparisons of calculations in the weak field and far field limits is the extent to which leading order quantum effects are local when the fields
are in the Boulware vacuum state. By ``local'' we mean that in the far field limit the leading order terms in the stress-energy tensor depend only on the leading order deviations of the metric from a flat space metric.  Comparison of the calculations in Ref.~\cite{abf} for Schwarzschild spacetime with the weak field calculations of Ref.~\cite{mazz} for nonrelativistic static spherically symmetric stars shows that while there is agreement in the leading order asymptotic behaviors of the energy density and pressure for massless scalar fields with conformal and minimal coupling to the scalar curvature, for all other couplings there are differences.  Further there are also disagreements in the leading order behavior of the quantity $\langle \phi^2 \rangle$ for all couplings to the scalar curvature except minimal coupling.
In Ref.~\cite{af} two of us computed the leading order difference in the stress-energy tensors in the far field limit for a Schwarzschild black hole and a static spherically symmetric star for massless scalar fields in the Boulware state.  It was shown that in most cases
the leading order behaviors of the quantities $\langle \phi^2 \rangle$ and $\langle T_{ab} \rangle$ have contributions from the local geometry as well as
nonlocal contributions from a zero frequency and zero angular momentum mode.  However, when the curvature coupling is minimal there is no difference at leading order between the case when a black hole is present and the case when a star is present.  When the coupling is conformal there is a difference for $\langle \phi^2 \rangle$, but for $\langle T_{ab} \rangle$ the differences cancel,  presumably because of the conformal symmetry.  In Ref.~\cite{mazz2} the calculation in Ref.~\cite{af} was generalized to the case of $D$ dimensions with $D \ge 4$.  As for $D = 4$, it was found that there is no leading order difference in $\langle \phi^2 \rangle$ and $\langle T_{ab} \rangle$ for minimal coupling and no leading order difference in $\langle T_{ab} \rangle$ for conformal coupling in $D$ dimensions.

In this paper, analytic solutions to the leading order mode equations in the far field limit of a large class of asymptotically flat, static, spherically symmetric spacetimes are found for both massless spin 0 and spin $1/2$ fields.
  From these solutions analytic expressions for $\langle \phi^2 \rangle$ for scalar fields and $\langle T_{ab} \rangle$ for both types of fields are derived when the fields are in the
  Boulware vacuum state.  For the spin $1/2$ field it
is shown that the leading order behavior of $\langle T_{ab} \rangle$ depends only on the geometry in the far field region.  For the conformally coupled scalar field it is shown that the leading order behavior of $\langle T_{ab} \rangle$ also depends only on the geometry in the far field region, but this is not true for $\langle \phi^2 \rangle$.  For the minimally coupled scalar field it is shown that there
are cases, such as spacetimes containing a star or a black hole, where the leading order behaviors of both $\langle \phi^2 \rangle$ and $\langle T_{ab} \rangle$ depend only on the geometry in the far field region.  However, for the minimally coupled scalar field in wormhole spacetimes the leading order behaviors of these quantities depend upon the entire geometry.

Numerical computations of the full stress-energy tensor  for the spin $1/2$ field in Schwarzschild, extreme Reissner-Nordstr\"{o}m, and three wormhole spacetimes are analyzed and displayed.
These provide important checks on the analytic calculations.  In each case the leading order behavior in the far field limit has been found to agree with the analytic results to within the numerical error of the computations.

In Section~\ref{sec:general-scalar} the computation of the stress-energy tensor for a massless scalar field is reviewed.  In Section~\ref{sec:WKB}, a somewhat peculiar method mentioned in~\cite{abf} of using the WKB approximation and conservation of the stress-energy tensor to compute the leading order behavior of the stress-energy tensor for massless scalar fields in
the far field limit of Schwarzschild spacetime is given.  Section~\ref{sec:general-spin-12} contains a review of the computation of the stress-energy tensor for the massless spin $1/2$ field.
      In Section~\ref{sec:Fermions} an analytic computation is made of the leading order behavior of the stress-energy tensor for the massless spin $1/2$ field in
    the far field limit.  A similar analytic computation is made for a massless scalar field with arbitrary coupling $\xi$ to the scalar curvature in Section~\ref{sec:Bosons}.
   In Section~\ref{sec:numeric} some of the numerical results for the massless spin $1/2$ field are displayed and discussed.  Section~\ref{sec:numsch} contains a comparison between
our results and previous analytic calculations of the leading order quantum corrections to the gravitational potential in Schwarzschild spacetime.
Some conclusions are given in Section~\ref{sec:conclusions}.  Various identities relating to modified Bessel functions which were used in the analytic computations are proven in the Appendix.
Throughout we use units such that $\hbar = c = G = 1$, and our conventions are those of Misner, Thorne, and Wheeler~\cite{mtw}.

\section{General Form for the Stress-Energy Tensor for Massless Scalar Fields}
\label{sec:general-scalar}

In this section the method developed to compute the stress-energy tensor for scalar fields in static spherically symmetric spacetimes given in
Ref.~\cite{ahs2} is adapted to the special case of computing the leading order components of this tensor for a class of asymptotically flat spacetimes in the far field limit
when the fields are in the Boulware state.

The metric for a general static spherically symmetric spacetime can be written as
\be ds^2 = - f(r) dt^2 + h(r) dr^2 + r^2 d \Omega^2 \;.  \label{metric} \ee
In this paper we consider the class of asymptotically flat spacetimes for which the metric functions in the large $r$ limit have the behaviors
\be \label{metriclimit}
     f(r) = 1- {2A \over r} + O\left( 1 \over r^2 \right) \qquad \hbox{and}  \qquad h(r) = 1 + {2B \over r} + O\left( 1 \over r^2 \right) \; ,
\ee
where $A$ and $B$ are two parameters describing the metric at large distances; for example, for both the Schwarzschild and Reissner-Nordstr\"om metrics, $A=B=M$.

In Ref.~\cite{ahs2} it was shown that the stress-energy tensor for a scalar field in a general static spherically symmetric spacetime can be written in terms of an analytic term plus a term involving sums and integrals over five different combinations of the modes, which must often be compute numerically,
\be
\langle T_{ab} \rangle = \langle T_{ab} \rangle_{\rm numeric} +  \langle T_{ab} \rangle_{\rm analytic} \;. \label{Tab-split}
\ee
For fields in the Boulware state, the quantity $\langle {T^a}_b \rangle_{\rm analytic}$ behaves as $r^{-6}$ at large $r$, but as we shall see,  $\langle {T^a}_b \rangle_{\rm numeric}$ dominates it and behaves as $r^{-5}$.
For a massless scalar field with coupling $\xi$ to the scalar curvature, $\langle {T^a}_b \rangle_{\rm numeric}$ has the asymptotic form\footnote{The complete forms for  $\langle T_{ab} \rangle_{\rm numeric}$ and the $S_n$ sums are given in Ref.~\cite{ahs2}. Here we present them only to leading order in the metric perturbations $A$ and $B$.}
\bes\label{TabSns}
\bea
      \langle {T^t}_t \rangle_{\rm numeric} &=& \left( 2 \xi + \frc12 \right) S_1 + \left( 2 \xi - \frc12 \right) S_2
                   + \left( 2 \xi - \frc12 \right) \frac{1}{r^2} S_3  - (2\xi-\frc12) {1 \over 4r^2} S_5  \;,  \\
      \langle {T^r}_r \rangle_{\rm numeric} &=& - \frc12 S_1 +  \frc12 S_2 - {1 \over 2 r^2} S_3  + \frac{2\xi }{r}  S_4 + {1 \over 8 r^2} S_5 \;,  \\
      \langle {T^\theta}_\theta \rangle_{\rm numeric} &=& \left( 2 \xi - \frc12 \right) S_1 + \left( 2 \xi - \frc12 \right) S_2 + {2 \xi \over r^2} S_3  - {\xi \over r} S_4 - {\xi \over 2 r^2} S_5 \; .
\eea
\ees
Here the $S_n$'s are sums and integrals over the radial modes $p_{\omega \ell}(r)$ and  $q_{\omega \ell}(r)$ of the Euclidean Green function.  In the Boulware vacuum state we anticipate that
to leading order the nonzero components of the stress-energy tensor in the asymptotically flat region go like $O(r^{-5})$.  Hence we need to determine $S_1$ and $S_2$ to $O(r^{-5})$, $S_3$ and $S_5$ to $O(r^{-3})$, and $S_4$ to $O(r^{-4})$.  To this order we find\footnote{Note that the dependence on the mode functions $p_{\omega \ell}$ and   $q_{\omega \ell}$ in these expressions is
exact.  It is the subtraction terms which are approximate.}
\bes\label{Sns}
\bea
    S_1 &=& \int_0^\infty {d\omega \over 4\pi^2} \omega^2 \left\{ \sum_{\ell = 0}^\infty \left[ (2\ell{+}1) p_{\omega \ell}(r)  q_{\omega \ell}(r)  - {1 \over r} - {A
        \over r^2} \right] +  \omega + {2A \omega \over r} \right\} \;, \label{S1def} \\
    S_2 &=& \int_0^\infty {d\omega \over 4\pi^2} \Biggl\{\sum_{\ell = 0}^\infty \left[ (2 \ell {+} 1) \frac{d p_{\omega \ell}(r)}{dr}  \frac{d q_{\omega \ell}(r)}{dr}
        + \left({1 \over 2r} + {2B + 3A \over 2r^2}\right) \omega^2 \right. \nn \\
    && \left. \quad {} + (\ell^2{+}\ell) \left({1 \over r^3} + {A + 2B \over r^4}\right) + {B-A \over 2r^4} \right]
        - \left(1 + {4A +2B \over r}\right){\omega^3 \over 3} + {A+B \over 3r^3} \omega \Biggr\} \;, \label{S2def} \\
    S_3 &=& \int_0^\infty {d\omega \over 4 \pi^2} \Biggl\{\sum_{\ell=0}^\infty \left[2(\ell{+}\frc12)^3 p_{\omega\ell}(r)q_{\omega\ell}(r) - {(\ell{+}\frc12)^2 \over r} -
        {(\ell{+}\frc12)^2 A\over r^2} + {\omega^2 r \over 2} + {3A \omega^2 \over 2} \right] \nn \\
    & & \qquad {}  - {2 r^2 \omega^3 \over 3} - {8Ar \omega^3 \over 3} + {\omega \over 4} + {A \omega \over 6r} - {B \omega\over 3r} \Biggr\} \;, \label{S3def} \\
    S_4 &=& \int_0^\infty {d\omega \over 4\pi^2} \left\{\sum_{\ell = 0}^\infty \left[ (2 \ell {+} 1) {d \over dr}\left[p_{\omega \ell}(r)  q_{\omega \ell}(r)\right]  +
        {1 \over r^2} + {2A \over r^3} \right]  - {2A \omega \over r^2} \right\} \;, \label{S4def} \\
    S_5 &=& \int_0^\infty {d\omega \over 4\pi^2} \left\{ \sum_{\ell = 0}^\infty \left[(2 \ell + 1) p_{\omega \ell}(r)  q_{\omega \ell}(r)  - {1 \over r} -
        {A \over r^2}\right] + \omega  + {2 A \omega \over r} \right\} \;. \label{S5def}
\eea
\ees
The same type of split also occurs for $\langle \phi^2 \rangle$ with
    \be \langle \phi^2 \rangle_{\rm numeric} = S_5 \;. \label{phi2num} \ee
For fields in the Boulware state, $\langle \phi^2 \rangle_{\rm analytic}$ goes like $r^{-4}$ at large $r$ so, as for the stress-energy tensor, $\langle \phi^2 \rangle_{\rm numeric}$  dominates.

The quantities $p_{\omega \ell}(r)$ and   $q_{\omega \ell}(r)$ are radial mode functions for the Euclidean Green function.  They satisfy the equation
\be\label{mode-eq-exact}
     {d^2S \over dr^2} + \left({2\over r} + {1 \over 2f} {df \over dr} - {1 \over 2h} {dh \over dr} \right) {dS \over dr} - \left[{\omega^2 h \over f} + {\left(\ell^2 + \ell\right) h \over r^2} + \xi R h \right] S = 0 \;,
\ee
where $R$ is the scalar curvature.  To first order in the metric parameters $A$ and $B$, this equation is
\be\label{mode-eq}
     {d^2S \over dr^2} + \left({2\over r} + {A+B \over r^2} \right) {dS \over dr} - \omega^2  \left(1+ {2A+2B \over r} \right) S
     - {\ell^2 + \ell \over r^2} \left(1 + {2B \over r} \right) S = 0 \; .
\ee
Also note that the relationships
\begin{subequations}
\bea\label{S4simple} S_4 &=& \frac{d S_5}{dr} \;, \\
\label{S2simple}
       S_2 &=& \frac{1}{2} \left[- {2h \over f} S_1 - {2h \over r^2} S_3 + \left({2 \over r} + {f' \over 2f} - {h' \over 2h} \right) S_4 + {dS_4 \over dr} - \left( 2 \xi R h - {h \over 2r^2} \right) S_5 \right]
       \eea
\end{subequations}
are exact.\footnote{To see this use the exact form of the mode sums in~\cite{ahs2} along with the mode equation~\eqref{mode-eq-exact}.}

The modes are normalized so that they satisfy the Wronskian condition
\be\label{wronskian}
     p_{\omega \ell}(r) \frac{d q_{\omega \ell}(r)}{dr} - \frac{d p_{\omega \ell}(r)}{dr} q_{\omega \ell}(r) = - \frac{1}{r^2} \left( \frac{h}{f} \right)^{1/2} \;.
\ee
In an asymptotically flat spacetime one boundary condition is that $q_{\omega \ell}(r)$ is finite in the limit $r \rightarrow \infty$.  The boundary
condition for $p_{\omega \ell}(r)$ depends on the behavior of the geometry at small values of $r$.  If the geometry is regular at $r=0$, such as in the case
of a static star, then $p_{\omega \ell}(r)$ is regular there as well.  If there is an event horizon then $p_{\omega \ell}(r)$ should be regular at the
event horizon.  For a wormhole $p_{\omega \ell}(r)$ should go to zero in the asymptotically flat region on the other side of the wormhole.

\section{WKB Approximation for the Stress-Energy Tensor for a Scalar Field}
\label{sec:WKB}

As shown in Ref.~\cite{ahs2}, for a scalar field the WKB approximation for the radial mode functions is obtained by first making the transformation
\bes\label{WKBansatz}
\bea
   p_{\omega \ell} &=& \frac{1}{(2 r^2 W)^{1/2}} \exp\left\{ \int^r W \left(\frac{h}{f} \right)^{1/2} dr \right\} \;, \\
   q_{\omega \ell} &=& \frac{1}{(2 r^2 W)^{1/2}} \exp\left\{ -\int^r W \left(\frac{h}{f} \right)^{1/2} dr \right\} \;.
\eea
\ees
Note that the Wronskian condition~\eqref{wronskian} is identically satisfied by these expressions.  Substitution into the mode equation \eqref{mode-eq-exact}
gives in both cases
\bea
W^2 &=& \Omega^2 + V_1 + V_2 + \frac{1}{2} \left[ \frac{f}{h W} \frac{d^2 W}{dr^2} + {d \over dr} \left(f \over h \right) \frac{1}{2 W} \frac{d W}{dr} - \frac{3}{2} \frac{f}{h} \left( \frac{1}{W} \frac{d W}{dr} \right)^2 \right] \;,
\label{W2eq}
\eea
with
\bes
\bea
\Omega^2 &=& \omega^2 + \left( \ell + \frac{1}{2} \right)^2 \frac{f}{r^2} \;,  \\
V_1 &=& \frac{1}{2 r h}\frac{df}{dr} - \frac{f}{2 r h^2} \frac{dh}{dr} - \frac{f}{4 r^2}  \;,  \\
V_2 &=& \xi f R \;.
\eea
\ees
Eq.~\eqref{W2eq} can be solved iteratively.  The zeroth order solution is $W = \Omega$.  The second order solution is
\be W = \Omega + \frac{1}{2 \Omega} (V_1 + V_2) - \frac{1}{8 \Omega^3} V_1^2 + \frac{1}{4} \left[ \frac{f}{h \Omega^2} \frac{d^2 \Omega}{dr^2}
            + {d \over dr} \left(f \over h \right) \frac{1}{2 \Omega^2} \frac{d \Omega}{dr}
            - \frac{3}{2} \frac{f}{h} \frac{1}{\Omega^3} \left( \frac{d \Omega}{dr} \right)^2 \right] \;.
\label{WKBapprox}
\ee

An approximation to $\langle T_{ab} \rangle_{\rm numeric}$ can be obtained
by first substituting Eqs.~\eqref{WKBansatz} into the sums in Eqs.~\eqref{Sns}.  Approximations for these sums, which we call
$(S_n)_{\rm WKBfin}$, can be obtained by substituting
the WKB expansion to some order for $W$ and its derivatives into the resulting expressions, keeping only terms which are smaller than or equal to the given
order of the WKB expansion.  Note that the order is increased by one for every radial derivative of $W$.  The details of how the  mode sums and
integrals are computed are given in Appendix F of Ref.~\cite{ahs2}.
The results are then substituted into Eqs.~\eqref{TabSns}
to obtain the approximation which we call $\langle T_{ab} \rangle_{\rm WKBfin}$.   While the exact expression for $\langle T_{ab} \rangle_{\rm numeric}$ is conserved, in general $\langle T_{ab} \rangle_{\rm WKBfin}$  is not a conserved tensor.

If a second order WKB expansion is used for Schwarzschild spacetime, where
     \be\label{Schwarz}  f = h^{-1} = 1 - 2 M/r \;,  \ee
then at lowest order there are terms in the expressions for $\langle {T^a}_b \rangle_{\rm WKBfin}$ which at large $r$ go like $1/r^4$,
$M/r^5$, $M^2/r^6$, and so forth.\footnote{Higher order terms in the WKB expansion also generate terms of order $1/r^4$, $M/r^5$, etc.;
however, the correct leading order behaviors for the sums $S_3$, $S_4$, and $S_5$ are obtained by using only a second order WKB expansion.}  The lowest order terms are present in the $M \rightarrow 0$ limit which is flat space.
Thus there is some justification in ignoring them since the full renormalized stress tensor is zero in flat space.  The terms of order
$M^2/r^6$ and higher are of subleading order at large $r$, so it is reasonable to ignore them.  That leaves the terms of order $M/r^5$.  Terms of this form do not appear in $\langle {T^a}_b \rangle_{\rm analytic}$,
which has instead leading order terms proportional to $M^2/r^6$.

The terms of order $M/r^5$ in the stress-energy tensor come from terms of order $M/r^3$ in
$(S_3)_{\rm WKBfin}$ and $(S_5)_{\rm WKBfin}$, $M/r^4$ in $(S_4)_{\rm WKBfin}$ and $M/r^5$ in $(S_1)_{\rm WKBfin}$ and $(S_2)_{\rm WKBfin}$.
We find that the terms of these orders
in $(S_3)_{\rm WKBfin}$, $(S_4)_{\rm WKBfin}$, and $(S_5)_{\rm WKBfin}$ are
\bes\label{S345}
\bea
  S_3 &=& -\frac{7 M}{480 \pi^2 r^3} \;,  \\
  S_4 &=& -\frac{M}{8 \pi^2 r^4} \;, \\
  S_5 &=& \frac{M}{24 \pi^2 r^3}  \;.
\eea
\ees
These are in agreement with both the numerical computations discussed in~\cite{abf} and the exact analytic computations shown in Section VI.
This is not the case for the corresponding terms in $(S_1)_{\rm WKBfin}$ and $(S_2)_{\rm WKBfin}$.
However, if Eqs.~\eqref{S345} along with expressions of the form $S_1 = a M/r^5$ and $S_2 = b M/r^5$, are substituted into Eqs.~\eqref{TabSns}, then the result is conserved to leading order if and only if
\bes
\bea
  a &=& \frac{3}{80 \pi^2} \;,  \\
  b &=& \frac{9}{80 \pi^2} \;.
\eea
\ees
The resulting expressions for the order $M/r^5$ contributions to $S_1$ and $S_2$ are in agreement with both our numerical and our exact analytical results.
If these results are substituted into Eq.~\eqref{TabSns} then one finds that the leading order behavior is
\bea
 \langle {T^a}_b \rangle = {M \over 40 \pi^2 r^5}\left[ {\rm diag}\left(-1,2,-3,-3\right) + 5 \xi {\rm diag} \left(2,-2,3,3\right) \right]\; , \label{Tmnsch}
\eea
which is the same as that given in Ref.~\cite{abf}.

Note that one might expect a similar calculation to work for the massless spin $1/2$ field.  However, because the second order WKB approximation, when used in the
way described above, does not give the correct leading order behaviors for all of the sums in Eqs.~\eqref{Sns}, we do not pursue this method further here.

\section{General Form of the Stress-Energy Tensor for the Massless Spin $1/2$ Field}
\label{sec:general-spin-12}

In this section the method developed to compute the stress-energy tensor for the massless spin $\frc12$ field in static spherically symmetric spacetimes given in
Ref.~\cite{gac} is adapted to the special case of computing the leading order components of this tensor for a class of asymptotically flat spacetimes in the far field limit.

In Ref.~\cite{gac} it was shown that the stress-energy tensor for a massless spin $\frc12$ field in a general static spherically symmetric spacetime can be written in the same way as Eq.~\eqref{Tab-split} with an analytic term plus a term involving sums and integrals over the modes.  As for the scalar field, $\langle {T^a}_b \rangle_{\rm analytic}$ gives a leading order contribution at
large $r$ of $1/r^6$ which is less than the $1/r^5$ contribution that we expect from $\langle {T^a}_b \rangle_{\rm numeric}$.  For a metric with the asymptotic form~\eqref{metriclimit}, $\langle {T^a}_b \rangle_{\rm numeric}$ has the asymptotic form\footnote{The complete form for  $\langle T_{ab} \rangle_{\rm numeric}$ is given in Ref.~\cite{gac}. Here we present it only to leading order in the metric perturbations $A$ and $B$.}
\bes
\bea
      \langle {T^t}_t \rangle_{\rm numeric} &=& \frac{1}{\pi^2} \int_0^\infty d \omega \left[ \left(\frac{1}{r^2} + \frac{2 A}{r^3} \right) \omega^2 A_1
         -\left(1 + \frac{4A}{r} \right) \omega^3 \right]  \; , \\
      \langle {T^r}_r \rangle_{\rm numeric} &=& \frac{1}{\pi^2} \int_0^\infty d \omega \left[ \left(\frac{1}{r^3} + \frac{A}{r^4} \right) \omega A_2
        - \left(\frac{1}{r^2} + \frac{2 A}{r^3} \right) \omega^2 A_1 + \left(\frac{1}{3} + \frac{4 A}{3r} \right) \omega^3
         + \frac{A+B}{6 r^3} \omega \right] \; ,  \nn \\ && \\
         \langle {T^\theta}_\theta \rangle_{\rm numeric} &=& \frac{1}{2\pi^2} \int_0^\infty d \omega \left[- \left(\frac{1}{r^3} + \frac{A}{r^4} \right) \omega A_2
          + \left(\frac{2}{3} + \frac{8 A}{3 r} \right) \omega^3 - \frac{A+B}{6 r^3} \omega \right] \; .
            \eea
\label{Tab12}
\ees
The asymptotic forms of the sums $A_1$ and $A_2$ that we use for the calculations in Section~\ref{sec:Fermions} are not simply the
asymptotic forms of the exact equations for $A_1$ and $A_2$ given in Ref.~\cite{gac}.  To eventually derive the form of the equations
which we use, it is useful to first discuss the exact expressions for $A_1$ and $A_2$ given in Ref.~\cite{gac}.  They are
\bes\label{A1A2}
\bea
     A_1 &=& \sum_{j=1/2}^\infty \left[ (j{+}\frc12) F^q_{\omega, j} F^p_{\omega,j} - (j{-}\frc12) G^q_{\omega,j-1} G^p_{\omega,j-1} + {r \over f^{1/2}} \right] \; , \label{A1}\\
     A_2 &=& \sum_{j=1/2}^\infty \left[ (j^2{-}\frc14) \left( G^q_{\omega,j} F^p_{\omega,j} + F^q_{\omega,j-1} G^p_{\omega,j-1} \right)
     - \frac{j^2{-}\frc14}{\omega} + \frac{r^2 \omega}{2f}\right] \;. \label{A2}
\eea
\ees
Here the sum is over half integer values of $j$, with $j \equiv \ell + \frc12$ in the notation of Ref.~\cite{gac}.  $F^p_{\omega, j}$ and $G^p_{\omega, j}$ are radial mode functions for the Euclidean Green function, as are $F^q_{\omega, j}$ and $G^q_{\omega, j}$.  The boundary conditions for the mode functions are the same as those for the scalar field.  The notation is also similar, so $F^q_{\omega, j}$ and $G^q_{\omega, j}$ are the modes which vanish at spatial infinity, while $F^p_{\omega, j}$ and $G^p_{\omega, j}$ are the modes which are regular at the origin if there is a star, or at the event horizon if there is a black hole.  They vanish at spatial infinity in the other universe if there is a wormhole.
Each pair satisfies the coupled set of equations
\bes\label{FGeqn}
\bea
    \frac{1}{h^{1/2}} {d \over dr} G_{\omega, j} &=& \frac{\omega}{f^{1/2}} F_{\omega, j} - \frac{j{+}\frc12}{r} G_{\omega, j} \; , \label{Geqn} \\
    \frac{1}{h^{1/2}} {d \over dr} F_{\omega, j} &=& \frac{\omega}{f^{1/2}} G_{\omega, j} + \frac{j{+}\frc12}{r} F_{\omega, j} \; . \label{Feqn}
\eea
\ees
The radial mode functions also satisfy the Wronskian condition
\be
     \omega [ G^q_{\omega,j} F^p_{\omega,j} - F^q_{\omega,j} G^p_{\omega,j}] = 1 \;. \label{Wronskianspin12}
\ee

Equations~\eqref{A1A2} can be recast in a more useful form in which the subscripts are consistently the same.  As a first step, rewrite each sum as the limit of a finite sum, and then shift $j \rightarrow j{+}1$ on those terms in the sum for which the subscript is $j{-}1$ rather than $j$.  The result is
\bes\label{A1A2shift}
\bea
     A_1 &=& \lim_{J \rightarrow \infty} \left\{ \sum_{j=1/2}^J \left[ (j{+}\frc12) F_{\omega,j}^q F^p_{\omega,j} + {r \over f^{1/2}} \right] - \sum_{j=-1/2}^{J-1} (j{+}\frc12) G^q_{\omega,j} G^p_{\omega,j} \right\}  \nn \\
     &=& \lim_{J \rightarrow \infty} \left\{ \sum_{j=1/2}^J \left[ (j{+}\frc12) \left(F_{\omega,j}^q F^p_{\omega,j} - G^q_{\omega,j} G^p_{\omega,j} \right) + {r \over f^{1/2}} \right] + (J{+}\frc12) G^q_{\omega,J} G^p_{\omega,J} \right\} \; ,  \label{A1shift} \\
     A_2 &=& \lim_{J \rightarrow \infty} \left\{ \sum_{j=1/2}^J \left[(j^2{-}\frc14) G^q_{\omega,j} F^p_{\omega,j} - \frac{j^2{-}\frc14}{\omega} + \frac{r^2 \omega}{2f}  \right] + \sum_{j=-1/2}^{J-1} (j^2 {+} 2j {+} \frc34) F^q_{\omega,j} G^p_{\omega,j} \right\} \nn \\
     &=& \lim_{J \rightarrow \infty} \left\{ \sum_{j=1/2}^J \left[2(j{+}\frc12)^2F^q_{\omega,j} G^p_{\omega,j} + \frac{r^2 \omega}{2f}  \right]- (J^2 {+} 2J {+} \frc34) F^q_{\omega,J} G^p_{\omega,J} \right\} \; . \label{A2shift}
\eea
\ees
Here the Wronskian condition~\eqref{Wronskianspin12} has been used in the final step to simplify the expression for $A_2$.

The limiting terms are to be evaluated in the large $J$ limit and for this purpose a WKB-like expansion can be used for the mode functions.  We first define new mode functions $Z^p_{\omega,j}$ and $Z^q_{\omega,j}$ as the ratio
     \be Z_{\omega,j} \equiv {F_{\omega,j} \over G_{\omega,j}} \; . \ee
Then Eqs.~\eqref{FGeqn} combine to become the single equation
\be\label{Zeqn}
    {1 \over h^{1/2}} {d \over dr} Z_{\omega,j} =  {\omega \over f^{1/2}} (1 - Z_{\omega,j}^2) + {2j{+}1 \over r} Z_{\omega,j}  \; .
\ee
Using the quadratic formula one finds the formal expression
\be\label{Zplusminus}
    Z_{\omega,j} = {(j{+}\frc12)f^{1/2} \over \omega r} \pm \left[ 1+ {(j{+}\frc12)^2 f \over \omega^2 r^2} - {f^{1/2} \over \omega h^{1/2}} {d \over dr} Z_{\omega,j} \right]^{1/2} \; .
\ee
This equation can be solved iteratively.  The zeroth order solution is obtained by setting $dZ_{\omega,j}/dr = 0$ on the right hand side.
 Since there are two solutions, one must correspond to $Z^q$ and the other to $Z^p$.  Because the $F^q$ and $G^q$ mode functions vanish at infinity, their magnitudes must decrease at large $r$.  From Eq.~\eqref{Feqn} it can be seen that the magnitude of $F^q$ can decrease as
 $r$ increases only if $G^q$ and $F^q$ have opposite signs.  Therefore
    \be\label{Gratiobound} Z^q_{\omega,j} =  F_{\omega,j}^q/G_{\omega,j}^q < 0  \ee
at large $r$.  Thus $Z^q$ must correspond
 to the solution with the minus sign, and therefore
$Z^p$ corresponds to the solution with the plus sign.
 In the large $j$ limit, we then find that
\bes\label{Zpqlimit}
\bea
    Z_{\omega,j}^p &=& {(2j{+}1)f^{1/2} \over \omega r} + O(1) \; , \\
    Z_{\omega,j}^q &=& - {\omega r \over (2j{+}1) f^{1/2}} + O(j^{-2}) \; .
\eea
\ees

Any product of $p$- and $q$-modes can be written, with the help of Eq.~\eqref{Wronskianspin12}, in terms of the $Z$'s; for example,
\bes
\bea
    G_{\omega,j}^q G_{\omega,j}^p &=& {1 \over \omega (Z_{\omega,j}^p - Z_{\omega,j}^q)} \; , \\
    F_{\omega,j}^q G_{\omega,j}^p &=& {Z_{\omega,j}^q \over \omega (Z_{\omega,j}^p - Z_{\omega,j}^q)} \; .
\eea
\ees
Substituting the approximate form~\eqref{Zpqlimit} gives
\bes\label{A1A2largeJ}
\bea
    \lim_{J \rightarrow \infty} (J{+}\frc12) G^q_{\omega,J} G^p_{\omega,J} &=& {r \over 2f^{1/2}} \; , \\
    \lim_{J \rightarrow \infty} (J^2 {+} 2J {+} \frc34) F^q_{\omega,J} G^p_{\omega,J} &=& - {r^2 \omega \over 4f} \; .
\eea
\ees
Substituting Eqs.~\eqref{A1A2largeJ} into Eqs.~\eqref{A1A2shift} then yields
\bes\label{A1A2better}
\bea
     A_1 &=& \sum_{j=1/2}^\infty \left[ (j{+}\frc12) \left(F_{\omega,j}^q F^p_{\omega,j} - G^q_{\omega,j} G^p_{\omega,j} \right) + {r \over f^{1/2}} \right] + {r \over 2f^{1/2}} \; , \\
     A_2 &=& \sum_{j=1/2}^\infty \left[2 (j{+}\frc12)^2 F^q_{\omega,j} G^p_{\omega,j} + \frac{r^2 \omega}{2f}\right] + {r^2 \omega \over 4f} \; .
\eea
\ees
Eqs.~\eqref{A1A2better} are completely general.  To first order in $A$ and $B$ they become
\bes\label{A1A2best}
\bea
     A_1 &=& \sum_{j=1/2}^\infty \left[(j{+}\frc12) \left(F_{\omega,j}^q F^p_{\omega,j} - G^q_{\omega,j} G^p_{\omega,j} \right) + r + A \right] + \frc12r + \frc12A \; , \\
     A_2 &=& \sum_{j=1/2}^\infty \left[2 (j{+}\frc12)^2 F^q_{\omega,j} G^p_{\omega,j} + \frc12r^2 \omega +Ar\omega \right] + \frc14r^2 \omega + \frc12Ar\omega \; .
\eea
\ees
It is this form which we use for our computations in Section~\ref{sec:Fermions}.

\section{Analytic Computation of the Stress-Energy Tensor for the Spin $\frc12$ Field}
\label{sec:Fermions}
In this section the asymptotic behavior of the stress-energy tensor for a massless spin $1/2$ field is computed analytically.  To do so we first use Eq.~\eqref{metriclimit} in Eqs.~\eqref{FGeqn} to write
in the large $r$ limit the approximate mode equations
\bes\label{FGmode2}
\bea
   {d F_{\omega,j} \over dr} - {j+\frc12 \over r}\left(1+{B \over r}\right) F_{\omega,j} &=& \omega \left(1 + {A+B \over r} \right) G_{\omega,j} \; , \\
   {d G_{\omega,j} \over dr} + {j+\frc12 \over r}\left(1+{B \over r}\right) G_{\omega,j} &=& \omega \left(1 + {A+B \over r} \right) F_{\omega,j} \; .
\eea
\ees
It is useful to change variables.  First let
\bes\label{FGtilde}
\bea
    \tilde F_{\omega,j} &\equiv& F_{\omega,j} + {j+\frc12 \over r}BF_{\omega,j} + \omega B G_{\omega,j} \; ,  \\
    \tilde G_{\omega,j} &\equiv& G_{\omega,j} - {j+\frc12 \over r}BG_{\omega,j} + \omega B F_{\omega,j} \; ,
\eea
\ees
and define
     \be\label{xdef} x\equiv\omega r \; . \ee
The mode equations are then
\bes\label{FGmode3}
\bea
    {d \tilde F_{\omega,j} \over dx} - {j+\frc12 \over x} \tilde F_{\omega,j} &=& \left(1+{C \omega \over x} \right) \tilde G_{\omega,j} \; ,  \\
    {d \tilde G_{\omega,j} \over dx} + {j+\frc12 \over x} \tilde G_{\omega,j} &=&  \left(1+{C \omega \over x} \right) \tilde F_{\omega,j} \; ,
\eea
\ees
with
    \be C \equiv A+B \; .\ee
The functions $\tilde F_{\omega,j}$ and $\tilde G_{\omega,j}$  satisfy, to first order in $j B/r$, the same Wronskian condition~\eqref{Wronskianspin12} that $F_{\omega,j}$ and $G_{\omega,j}$ do.

If $C=0$, it is not hard to show that the solutions to Eqs.~(\ref{FGmode3}) will be given by $(x/\omega)^{1/2}I_j(x)$ and $(x/\omega)^{1/2}K_j(x)$ for $\tilde F_{\omega,j}$ and  $(x/\omega)^{1/2}I_{j+1}(x)$ and $(x/\omega)^{1/2}K_{j+1}(x)$ for $\tilde G_{\omega,j}$, where $I_j$ and $K_j$ are modified Bessel functions.  Many of the properties of these functions, including a number of useful identities, can be found in Appendix~\ref{bessel}.  The $I_j$'s are well behaved at small $x$, but diverge at large $x$, while the $K_j$'s are infinite at $x=0$, but damp to zero in the limit $x \rightarrow \infty$.  Recalling that $x = \omega r$, we thus anticipate that the  $F^q$ and $G^q$ modes will primarily involve the $K_j$'s.

For $C \ne 0$ we can continue to write $\tilde F_{\omega,j}$ and $\tilde G_{\omega,j}$ in terms of $(x/\omega)^{1/2}I_j(x)$ and $(x/\omega)^{1/2}K_j(x)$ by defining the new pairs of variables
$\alpha^p_{\omega,j}(x)$,  $\beta^p_{\omega,j}(x)$ and $\alpha^q_{\omega,j}(x)$,  $\beta^q_{\omega,j}(x)$  such that for each pair
\bes\label{defalphabeta}
\bea
    \tilde F_{\omega,j}(x) &=& \left(\frac{x}{\omega}\right)^{1/2} \left[\alpha_{\omega,j}(x) I_j(x) - \beta_{\omega,j}(x) K_j(x) \right] \; , \label{Falphabeta}  \\
    \tilde G_{\omega,j}(x) &=& \left(\frac{x}{\omega}\right)^{1/2} \left[\alpha_{\omega,j}(x) I_{j+1} (x) + \beta_{\omega,j}(x) K_{j+1}(x) \right] \; . \label{Galphabeta}
\eea
\ees
Using the Wronskian condition~(\ref{Wronsk2}) for the Bessel functions gives to first order in $j B/r$ the condition
     \be\label{alphabetaconstraint} \alpha^p_{\omega,j}(x)\beta^q_{\omega,j}(x) - \alpha^q_{\omega,j}(x)\beta^p_{\omega,j}(x) = 1  \; .\ee
Substituting Eqs.~(\ref{defalphabeta}) into the mode equations (\ref{FGmode3}) and using the recursion relations~\eqref{recursion} to eliminate the derivatives acting on the modified Bessel functions, yields coupled linear expressions for $\alpha_{\omega,j}^\prime(x)$ and $\beta_{\omega,j}^\prime(x)$.  These can then be simplified using Eq.~(\ref{Wronsk2}) to give
\bes\label{alphabetamode}
\bea
    \alpha_{\omega,j}^\prime(x) &=& C\omega \left\{\alpha_{\omega, j}(x) \left[I_{j+1}(x) K_{j+1}(x) + I_j(x)K_j(x)\right] + \beta_{\omega,j}(x) \left[K_{j+1}^2(x) - K_j^2(x) \right] \right\} \; ,  \quad \\
    \beta_{\omega, j}^{\, \prime}(x) &=& C\omega \left\{\alpha_{\omega,j}(x) \left[I_j^2(x) - I_{j+1}^2(x)\right] - \beta_{\omega,j}(x) \left[I_{j+1}(x) K_{j+1}(x) + I_j(x)K_j(x)\right] \right\} \; ,
\eea
\ees
where primes denote derivatives with respect to $x$.

To compute the stress-energy tensor, we shall require the products of the $p$- and $q$-solutions, and for this reason it is helpful to work with the following equations for the relevant products of the $\alpha$'s and $\beta$'s:
\bes\label{productmode}
\bea
    \left[\alpha_{\omega,j}^p(x) \alpha_{\omega,j}^q(x)\right]^\prime &=& C \omega \Bigl\{ 2 \alpha_{\omega,j}^p(x) \alpha_{\omega,j}^q(x) \left[I_{j+1}(x)K_{j+1}(x) + I_j(x) K_j(x) \right] \nn \\
    &&\qquad {} + \left[\alpha_{\omega,j}^p(x) \beta_{\omega,j}^q(x) + \beta_{\omega,j}^p(x) \alpha_{\omega,j}^q(x) \right] \left[K_{j+1}^2(x) - K_j^2(x) \right] \Bigr\} \; , \\
    \left[\alpha_{\omega,j}^p(x) \beta_{\omega,j}^q(x)\right]^\prime &=& C \omega \Bigl\{ \beta_{\omega,j}^p(x) \beta_{\omega,j}^q(x) \left[K_{j+1}^2(x) - K_j^2(x) \right] \nn \\
    &&\qquad {} + \alpha_{\omega,j}^p(x) \alpha_{\omega,j}^q(x)  \left[I_j^2(x) - I_{j+1}^2(x) \right] \Bigr\} \; , \\
    \left[\beta_{\omega,j}^p(x) \alpha_{\omega,j}^q(x)\right]^\prime &=& C \omega \Bigl\{ \beta_{\omega,j}^p(x) \beta_{\omega,j}^q(x) \left[K_{j+1}^2(x) - K_j^2(x) \right] \nn \\
    &&\qquad {}  + \alpha_{\omega,j}^p(x) \alpha_{\omega,j}^q(x)  \left[I_j^2(x) - I_{j+1}^2(x) \right] \Bigr\} \; , \\
     \left[\beta_{\omega,j}^p(x) \beta_{\omega,j}^q(x)\right]^\prime &=& C \omega \Bigl\{ \left[\alpha_{\omega,j}^p(x) \beta_{\omega,j}^q(x) + \beta_{\omega,j}^p(x) \alpha_{\omega,j}^q(x)\right] \left[I_j^2(x)-I_{j+1}^2(x) \right] \nn \\
    &&\qquad {} -2\beta_{\omega,j}^p(x) \beta_{\omega,j}^q(x) \left[I_{j+1}(x)K_{j+1}(x) + I_j(x) K_j(x) \right] \Bigr\} \; .
\eea
\ees
We want to solve these equations to first order in $C \omega$.  The solutions will be substituted into Eqs.~\eqref{A1A2best} and the sums over $j$ will be computed.  Then the
results will be substituted into Eqs.~\eqref{Tab12} and the integrals over $\omega$ will be computed.
For large values of $j$ the condition~\eqref{alphabetaconstraint} breaks down and for large enough values of $\omega$ the quantity $C \omega$ is not small. Therefore one
might be concerned that the leading order behavior of the stress-energy tensor will not be correct.  However, to leading order the nonzero components of the stress-energy tensor go like $D/r^5$ for some constant
$D$.  If we temporarily consider units in which $\hbar = c = 1$ but $G \ne 1$, then both $C$ and $D$ have units of length.  Thus $D$ can be proportional to $C$ but it cannot be proportional to $C^2$ or any other power of $C$.  For this reason it should be sufficient to only keep terms to order $C$ in both the condition~\eqref{alphabetaconstraint} and the solutions of Eqs.~\eqref{productmode}.

In flat space the only nonzero functions are $\alpha_{\omega,j}^p(x)$ and $\beta_{\omega,j}^q(x)$, and these are constants.
For the class of spacetimes we are considering we expect that this will continue to be true to leading order in a perturbative expansion.  However, we cannot rule
out the possibility that pathological metrics exist for which this condition would be violated.  Thus our calculations may not cover every possible spacetime with an
asymptotic metric of the form~\eqref{metriclimit}, but as can be seen in Sec.~\ref{sec:numeric}, they definitely apply to two standard black hole metrics and some wormhole metrics and
there is every reason to believe that they apply more generally to most or all static spherically symmetric black holes, static stars, and wormholes.

 Hence, to leading order, we assume that on the right hand side of Eq.~(\ref{productmode}) all products other than $\alpha_{\omega,j}^p(x)\beta_{\omega,j}^q(x)$ vanish.
 Furthermore, the constraint~(\ref{alphabetaconstraint}) demands that to this order $\alpha_{\omega,j}^p(x)\beta_{\omega,j}^q(x) = 1$, so Eqs.~(\ref{productmode}) simplify to
\bes\label{productmode2}
\bea    \left[\alpha_{\omega,j}^p(x) \alpha_{\omega,j}^q(x)\right]^\prime &=& C \omega  \left[K_{j+1}^2(x) - K_j^2(x) \right] \; , \\
    \left[\alpha_{\omega,j}^p(x) \beta_{\omega,j}^q(x)\right]^\prime &=& 0  \;,  \label{apbqeq} \\
      \left[\beta_{\omega,j}^p(x) \alpha_{\omega,j}^q(x)\right]^\prime &=& 0  \; ,  \\
    \left[\beta_{\omega,j}^p(x) \beta_{\omega,j}^q(x)\right]^\prime &=& C \omega  \left[I_j^2(x)-I_{j+1}^2(x) \right]  \; .
\eea
\ees
We now need to find suitable boundary conditions for Eqs.~(\ref{productmode2}).  This can be accomplished in part by looking at the solutions to Eqs.~\eqref{alphabetamode} in the limit $x \rightarrow \infty$.  Using the usual asymptotic expansions for the Bessel functions one can show that for arbitrarily large values of $x$
\bes
\bea
I_{j+1}(x) K_{j+1}(x) + I_j(x) K_j(x) &\approx& \sum_{n=0} c_n \, x^{-n-1}  \; ,  \\
K_{j+1}^2(x) - K_j^2(x) &\approx& e^{-2x} \sum_{n=0}  d_n \, x^{-n-2}  \; ,  \\
I_{j}^2(x) - I_{j+1}^2(x) &\approx& e^{2x} \sum_{n=0} e_n \, x^{-n-2}  \; .
\eea
\ees
Note that $c_0 = 1$.  Asymptotic solutions to Eqs.~\eqref{alphabetamode} can be made by substituting the above expansions into these equations.  It turns out that
the following pairs of sums result in solutions to Eqs.~\eqref{alphabetamode} in the limit $x \rightarrow \infty$:
\bes
\bea
\alpha^p_{\omega,j}(x) &=&  \sum_{m = 0} a^p_m \, x^{C \omega -m}  \; ,  \\
 \beta^p_{\omega,j}(x) &=& e^{2 x} \sum_{m=0} b^p_m \, x^{C \omega -m -2} \; ,
\eea
\ees
and
\bes
\bea
\alpha^q_{\omega,j}(x) &=& e^{-2x} \sum_{m = 0}  a^q_m \, x^{-C \omega -m -2} \; , \\
 \beta^q_{\omega,j}(x) &=& \sum_{m=0} b^q_m \,x^{-C \omega -m} \;.
\eea
\ees
It is easily shown that $a^p_0$ and $b^q_0$ are both arbitrary constants whose values determine the values of the remaining coefficients.
Note that we have chosen the solutions for $\alpha^q_{\omega,j}$ and $ \beta^q_{\omega,j}$ so that the mode functions
$F^q_{\omega,j}$ and $G^q_{\omega,j}$ are well behaved in the limit $r \rightarrow \infty$, which is the correct boundary condition for these modes.

The boundary conditions for most of the pairs of sums are now easily obtained.  First note that
\bes
\bea
    \lim_{x \rightarrow \infty} \alpha^p_{\omega,j}(x) \alpha^q_{\omega,j}(x) &=& a_0^p \, a_0^q \lim_{x \rightarrow \infty} x^{-2} e^{-2x} = 0 \;, \label{bdya} \\
    \lim_{x \rightarrow \infty} \alpha^p_{\omega,j}(x) \beta^q_{\omega,j}(x) &=& a_0^p \, b_0^q  \;, \label{bdyb} \\
    \lim_{x \rightarrow \infty} \alpha^q_{\omega,j}(x) \beta^p_{\omega,j}(x)  &=& a_0^q \, b_0^p \lim_{x \rightarrow \infty} x^{-4} = 0 \;. \label{bdyc}
\eea
The Wronskian condition~\eqref{alphabetaconstraint} then gives
\be
    \lim_{x \rightarrow \infty} \alpha^p_{\omega,j}(x) \beta^q_{\omega,j}(x) = a_0^p \, b_0^q =  1 \;. \label{bdyd}
\ee
\ees
Note that all that is uniquely fixed is the product of the two functions in this limit.  However, this is all that is necessary for the computation of $\langle T_{ab} \rangle$.
Solving Eq.~\eqref{apbqeq} using \eqref{bdyd} one finds that to first order in $C\omega$ that
\be
  \alpha^p_{\omega,j}(x) \beta^q_{\omega,j}(x) = 1  \label{apbqfsol}
\ee
for all values of $x$.

To obtain a final boundary condition, we must impose the constraint that the $p$-solution be well-behaved at the inner boundary.  This is difficult because we do not in general have knowledge of the metric in this region.  We anticipate that in the region $r \sim {\rm max(|A|,|B|)}$, the metric will vary significantly from its large $r$ approximation~\eqref{metriclimit}.  Let us choose a distance $a$  which is sufficiently larger than ${\rm max(|A|,|B|)}$ so that the approximation~\eqref{metriclimit} can be assumed accurate for $r \stackrel{_>}{_\sim} a$, but which is small
  enough that $a \ll r_e$, with $r_e$ the value of $r$ at which we wish to evaluate the stress-energy tensor.  Because we remain agnostic about the metric for $r<a$, we cannot know the relative size of the $K$ and $I$ contribution to the $p$-modes at $r=a$. Looking at Eq.~\eqref{Galphabeta}, one might expect that generically
\be
\label{smallrfermi}
    \alpha_{\omega,j}^p(\omega a) I_{j+1}( \omega a) \sim \beta_{\omega, j}^p (\omega a)  K_{j+1} (\omega a) \; .
\ee
We anticipate that the most important contributions to the stress-energy tensor at the radius $r_e$ come from $\omega \sim r_e^{-1}$. Since $a \ll r_e$, we therefore expect that $\omega a$ will be small.  At small $x$, $K_j(x)$ and $I_j(x)$ behave as $x^{-j}$ and $x^j$ respectively.  Using Eq.~\eqref{smallrfermi}, it follows that
\be\label{smallfermi2}
\beta_{\omega,j}^p (\omega a) \sim (\omega a)^{2j+2} \alpha_{\omega,j}^p(\omega a) \; .
\ee
Multiplying both sides by $\beta_{\omega,j}^q(\omega a)$ and using Eq.~\eqref{apbqfsol}, we then have
\be\label{smallfermi3}
\beta_{\omega,j}^p (\omega a) \beta_{\omega,j}^q(\omega a) \sim (\omega a)^{2j+2} \alpha_{\omega,j}^p(\omega a) \beta_{\omega,j}^q(\omega a) \sim {a^{2j+2} \over r_e^{2j+2}} \; .
\ee
Note that the smallest $j$ we are interested in is $j=\frc12$, and hence Eq.~\eqref{smallfermi3} falls at least as fast as $r_e^{-3}$.  Recall that we are only interested in keeping terms to order $\omega C \sim C/r_e$.  For sufficiently large $r_e$, Eq.~\eqref{smallfermi3} will always be negligible compared to $C/r_e$, and hence we can treat this as zero, so that\footnote{It is conceivable that for certain pathological spacetime metrics this relationship will not hold, and hence the $\beta_j^p$ will be important.  That is why our argument is not completely general.}
\be\label{bdye}
    \lim_{x \rightarrow 0} \beta_{\omega,j}^p(x)\beta_{\omega,j}^q(x) = 0 \; .
\ee

Solving Eqs.~\eqref{productmode2} subject to the boundary conditions Eqs.~\eqref{bdya}, \eqref{bdyc}, \eqref{bdyd}, and \eqref{bdye} is now straightforward:
\bes\label{abproducts}
\bea
    \alpha_{\omega,j}^p(x) \alpha_{\omega,j}^q(x) &=& C\omega \int_x^\infty \left[K_j^2(y) - K_{j+1}^2(y)\right] dy \; , \\
    \alpha_{\omega,j}^p(x) \beta_{\omega,j}^q(x) &=& 1 \; , \\
    \beta_{\omega,j}^p(x) \alpha_{\omega,j}^q(x) &=& 0 \; , \\
    \beta_{\omega,j}^p(x) \beta_{\omega,j}^q(x) &=& C\omega \int_0^x \left[I_j^2(y) - I_{j+1}^2(y)\right] dy \; .
\eea
\ees

To compute the quantities $A_1$ and $A_2$,
 first solve Eqs.~(\ref{FGtilde}) for $F_{\omega,j}$ and $G_{\omega,j}$ in terms of $\tilde F_{\omega,j}$ and $\tilde G_{\omega,j}$ and then substitute into Eqs.~\eqref{A1A2best}.
  Next use Eqs.~(\ref{defalphabeta}) to find expressions for $A_1$ and $A_2$ in terms of $\alpha_j$ and $\beta_j$.  Then use the explicit forms for the products in Eqs.~(\ref{abproducts}).  Expanding to linear order in $C$ we find
\bes\label{A1A2expand}
\bea
    A_1 &=& \sum_{j=\frc12}^\infty \biggl[(j{+}\frc12)\biggl\{ Cx \left[I_j^2(x)-I_{j+1}^2(x)\right] \int_x^\infty \left[K_j^2(y) - K_{j+1}^2(y)\right]dy  \nn \\
    &&{} + Cx \left[K_j^2(x) - K_{j+1}^2(x) \right] \int_0^x \left[I_j^2(y) - I_{j+1}^2(y) \right] dy - \frac{x}{\omega}\left[I_j(x)K_j(x)+I_{j+1}(x)K_{j+1}(x)\right] \nn \\
    &&{} + B(2j+1)\left[I_j(x)K_j(x)-I_{j+1}(x)K_{j+1}(x) \right] \biggr\} + \frac{x}{\omega} + A\biggr] + \frac{x}{2\omega} + \frc12A \;,  \\
    A_2 &=& \sum_{j=\frc12}^\infty \biggl[2x(j{+}\frc12)^2 \biggl\{C I_j(x) I_{j+1}(x) \int_x^\infty \left[K_j^2(y) - K_{j+1}^2(y)\right] dy \nn \\
    && {} - C K_j(x)K_{j+1}(x) \int_0^x \left[I_j^2(y) - I_{j+1}^2(y) \right] dy - \frac{1}{\omega} I_{j+1}(x) K_j(x) \nn \\
    && {} + B \left[I_j(x)K_j(x) - I_{j+1}(x)K_{j+1}(x) \right]  \biggr\} + \frac{x^2}{2\omega} + Ax \biggr] + \frac{x^2}{4\omega} + \frc12Ax \; .
\eea
\ees

The sums over $j$ for each term in $A_1$ and $A_2$ have been computed in the Appendix.
The integral terms in the  formula for $A_1$ are given by Eq.~(\ref{identB}), while the other terms in the sum are given by Eqs.~(\ref{sum1d}) and~(\ref{sum1c}).  The integral terms in the
 formula for  $A_2$ are given by Eq.~(\ref{identC}), while the other terms in the sum are given by Eqs.~(\ref{sum2a}) and~(\ref{sum1c}).  The result is
\bes\label{A1A2simple}
\bea
    A_1 &=& C\left(2x^2\int_x^\infty {dy\over y} e^{-2y} +2x-\frc12-xe^{-2x}+\frc12e^{-2x} \right) -\frac{x}{\omega} (\frc12-x)+2B(\frc14-x) + \frac{x}{2\omega} + \frc12A \; ,  \nn \\ && \\
    A_2 &=& C\left(\frc43 x^3 \int_x^\infty {dy\over y}e^{-2y} -\frc23x^2e^{-2x}+\frc13xe^{-2x}+\frc16e^{-2x} +2x^2-\frc12x-\frc16 \right) - \frac{1}{\omega} (\frc14x^2-\frc23x^3) \nn \\
    &&\qquad {} + 2Bx(\frc14-x) + \frac{x^2}{4\omega} + \frc12Ax \; .
\eea
\ees
Simplifying these two expressions, and replacing $\omega$ with $x/r$, we find
\bes\label{A1A2simpler}
\bea
    A_1 &=& xr + 2Ax - Cx e^{-2x}  +\frc12Ce^{-2x} +2Cx^2\int_x^\infty {dy\over y} e^{-2y} \; ,  \\
    A_2 &=& \frc23 x^2 r + 2Ax^2 - \frc16C - \frc23C x^2e^{-2x} + \frc13C x e^{-2x} + \frc16Ce^{-2x} + \frc43C x^3 \int_x^\infty {dy\over y}e^{-2y} \; . \quad
\eea
\ees

To complete the computation of $\langle {T^a}_b \rangle$, first recall that the analytic contribution falls as $r^{-6}$ at large $r$, and hence can be neglected.  The numeric contribution, Eqs.~(\ref{Tab12}), is thus the dominant contribution.
Substituting Eqs.~(\ref{A1A2simpler}) into Eqs.~\eqref{Tab12} and changing the integration variable from $\omega$ to $x = \omega r$ gives to linear order in $C/r$
\bes\label{Tmunufermi2}
\bea
    \langle {T^t}_t \rangle &=& {C \over\pi^2r^5} \int_0^\infty \left[\frc12x^2e^{-2x}-x^3e^{-2x}+2x^4\int_x^\infty {dy \over y} e^{-2y} \right] dx \; , \\
    \langle {T^r}_r \rangle &=& {C \over\pi^2r^5} \int_0^\infty \left[\frc13x^3e^{-2x}-\frc16x^2e^{-2x} dx + \frc16xe^{-2x} - \frc23x^4\int_x^\infty {dy \over y} e^{-2y} \right] dx \; , \qquad \\
    \langle {T^\theta}_\theta \rangle = \langle {T^\phi}_\phi \rangle &=& {C \over 2 \pi^2 r^5} \int_0^\infty \left[ \frc23x^3e^{-2x} - \frc13x^2e^{-2x} - \frc16xe^{-2x} - \frc43 x^4 \int_x^\infty {dy\over y}e^{-2y}\right] dx \; .
\eea
\ees
Integrating the last term in each expression by parts we find that in the $(t,r,\theta,\phi)$ basis,
     \be\label{Tmunufermi3} \langle {T^a}_b \rangle = {C \over80\pi^2r^5} \, {\rm diag}\left(4,2,-3,-3\right) \; . \ee

\section{Analytic Computation of the Stress-Energy Tensor for Scalar Fields}
\label{sec:Bosons}
We now proceed to a computation of the quantities $\langle \phi^2 \rangle$ and $\langle T_{ab} \rangle$ for massless scalar fields.
As can be seen from Eqs.~\eqref{TabSns}, \eqref{Sns}, and~\eqref{phi2num} these quantities contain sums and integrals over the radial mode functions.
Thus we begin by finding approximate solutions to the mode equation~\eqref{mode-eq} which are valid at large values of $r$.

To solve the radial mode equation it is useful to first change variables in such a way that the modes and the equations they
satisfy are similar to those which are solved for the spin $\frc12$ field.  We begin by defining
 a new mode function $F$ such that
 \begin{subequations}
\be
     S \equiv {1 \over r^{1/2}} \left(F + {A\over 2r}F - B {dF \over dr} \right) \; . \label{F-def-inverse}
\ee
To first order in $A$ and $B$ the inverse relation is
\be\label{F-def}
     F = r^{1/2} \left(S+{B-A \over 2r}S + B{dS \over dr} \right) \;.
\ee
\end{subequations}
If~\eqref{F-def-inverse} is substituted into the mode equation~\eqref{mode-eq} then the resulting equation has terms
of zeroth, first, and second order in $A$ and $B$.  Since we are working to first order, the second order terms can
be dropped. One of the first order terms is proportional to $B d^3 F/dr^3$.  This term can be eliminated.  To do so
first take one derivative with respect to $r$ of the zeroth order equation and solve it for $d^3 F/dr^3$.  The result
is
\be
\frac{d^3 F}{d r^3} = - \frac{1}{2r} \frac{d^2 F}{d r^2} + \left( \omega^2 + \frac{(\ell + \frc12)^2}{r^2} + \frac{3}{2 r^2} \right) \frac{d F}{d r}
                          - \left(\frac{\omega^2}{2 r} +  \frac{5(\ell + \frc12)^2}{2 r^3}  \right) F  \;.  \label{F3p}
\ee
Substituting this into the equation for $F$ which contains both zeroth and first order terms, one finds that to first order $F$ satisfies the equation
\be
     {d^2F \over dr^2} + {1 \over r} {dF \over dr} - \left[\omega^2 + {2C \over r} \omega^2 + {(\ell+\frc12)^2 \over r^2} \right] F = 0 \; , \label{Feq}
\ee
where, as before, $C = A+B$.

There are two solutions to this equation, $F^p_{\omega,\ell}$ and $F^q_{\omega,\ell}$, for each value of $\omega$ and $\ell$.
 The Wronskian condition~(\ref{wronskian}) becomes
    \be {dF^q \over dr} F^p - {dF^p \over dr} F^q = - {1 \over r} \; . \label{Fwronskian} \ee
Defining $x=\omega r$ as before, and letting primes denote derivatives with respect to $x$, one finds that Eqs.~\eqref{Feq} and~\eqref{Fwronskian} become
\be\label{F-mode-eq}
     F^{\prime\prime}+ {1 \over x} F^\prime - \left[1 + {2C\omega \over x}  + {(\ell+\frc12)^2 \over x^2} \right] F = 0 \; ,
\ee
and
\be\label{WronskF} F^{q\prime} F^p - F^{p\prime} F^q = - {1 \over x} \; . \ee

If $C=0$, then the solutions to Eq.~(\ref{F-mode-eq}) are again modified Bessel functions.  As for the spin $\frc12$ field, we define new functions $\alpha_j$ and $\beta_j$, where $j = \ell + \frc12$, so that
\bes\label{alphabetadef}
\bea
     F(x) &=& \alpha_{\omega,j}(x) I_j(x) + \beta_{\omega,j}(x) K_j(x) \; , \label{noprimedef} \\
     F^\prime(x) &=& \alpha_{\omega,j}(x) I^\prime_j(x) + \beta_{\omega,j}(x) K^\prime_j(x) \; .
\eea
\ees
In general, this can always be done, since we can treat this as two equations in two unknowns.  Demanding that these equations be compatible implies
\be\label{ab-relate-1} \alpha^\prime_{\omega,j}(x) I_j(x) + \beta^\prime_{\omega,j}(x) K_j(x) = 0 \; . \ee
Substituting Eqs.~(\ref{alphabetadef}) into Eq.~(\ref{F-mode-eq}), and using the fact that the Bessel functions satisfy the modified Bessel's equation (\ref{besseldef}), we find
\be\label{ab-relate-2}
     \alpha^\prime_{\omega,j}(x) I^\prime_j(x) + \beta^\prime_{\omega,j}(x) K^\prime_j(x) = {2C\omega \over x} \left[ \alpha_{\omega,j}(x) I_j(x) + \beta_{\omega,j}(x) K_j(x) \right] \; .
\ee
Solving Eqs.~(\ref{ab-relate-1}) and (\ref{ab-relate-2}) for $\alpha^\prime_{\omega,j}$ and $\beta^\prime_{\omega,j}$, and using the modified Bessel function Wronskian Eq.~(\ref{Wronsk1}), one finds
\bes\label{abprime}
\bea
     \alpha^\prime_{\omega,j}(x) &=& 2C\omega \left[\alpha_{\omega,j}(x) K_j(x) I_j(x) + \beta_{\omega,j}(x) K_j^2(x) \right] \; , \\
     \beta^\prime_{\omega,j}(x) &=& -2C\omega \left[\alpha_{\omega,j}(x) I_j^2(x) + \beta_{\omega,j}(x) K_j(x) I_j(x) \right] \; .
\eea
\ees
The Wronskian~\eqref{WronskF} in terms of the $\alpha$'s and $\beta$'s is identical to leading order to Eq.~(\ref{alphabetaconstraint}):
     \be\label{alphabetaconstraints} \alpha_{\omega,j}^p(x)\beta_{\omega,j}^q(x) - \alpha_{\omega,j}^q(x)\beta_{\omega,j}^p(x) = 1 \; . \ee
As in Section~\ref{sec:Fermions}, we will work not with Eqs.~\eqref{abprime}, but with the derivatives of the corresponding products.  We find
\bes\label{scalarmodeproducts}
\bea
     \left[ \alpha_{\omega,j}^p(x) \alpha_{\omega,j}^q(x) \right]^\prime &=& 2C\omega \Bigl\{2\alpha_{\omega,j}^p(x) \alpha_{\omega,j}^q(x) K_j(x) I_j(x) \nn \\
     && \qquad {} + \left[\alpha_{\omega,j}^p(x) \beta_{\omega,j}^q(x) + \beta_{\omega,j}^p(x) \alpha_{\omega,j}^q(x) \right] K_j^2(x) \Bigr\} \; , \\
     \left[ \alpha_{\omega,j}^p(x) \beta_{\omega,j}^q(x) \right]^\prime &=& 2C\omega \left[ \beta_{\omega,j}^p(x) \beta_{\omega,j}^q(x) K_j^2(x) - \alpha_{\omega,j}^p(x) \alpha_{\omega,j}^q(x) I_j^2(x) \right] \; , \\
     \left[ \beta_{\omega,j}^p(x) \alpha_{\omega,j}^q(x) \right]^\prime &=& 2C\omega \left[ \beta_{\omega,j}^p(x) \beta_{\omega,j}^q(x) K_j^2(x) - \alpha_{\omega,j}^p(x) \alpha_{\omega,j}^q(x) I_j^2(x) \right] \; , \\
     \left[ \beta_{\omega,j}^p(x) \beta_{\omega,j}^q(x) \right]^\prime &=& - 2C\omega \Bigl\{ \left[\alpha_{\omega,j}^p(x) \beta_{\omega,j}^q(x) + \beta_{\omega,j}^p(x) \alpha_{\omega,j}^q(x) \right] I_j^2(x) \nn \\
     && \qquad {} + 2 \beta_{\omega,j}^p(x) \beta_{\omega,j}^q(x) K_j(x) I_j(x) \Bigr\} \; .
\eea
\ees
As for the spin $\frc12$ field, we keep only terms to first order in $C\omega$ in these equations.  Thus we need to only keep the variables on the right side to zeroth order.  As before, we expect that all of the products except for $\alpha_{\omega,j}^p(x) \beta_{\omega,j}^q(x)$ will vanish to leading order, and because of the Wronskian condition Eq.~\eqref{alphabetaconstraints}, we expect this product to be one to leading order.  Hence Eqs.~\eqref{scalarmodeproducts} can be simplified to
\bes\label{scalarmodeproducts2}
\bea
     \left[ \alpha_{\omega,j}^p(x) \alpha_{\omega,j}^q(x) \right]^\prime &=& 2C\omega K_j^2(x)  \; , \label{apaqseq} \\
     \left[ \alpha_{\omega,j}^p(x) \beta_{\omega,j}^q(x) \right]^\prime &=& 0 \;, \label{apbqseq} \\
     \left[ \beta_{\omega,j}^p(x) \alpha_{\omega,j}^q(x) \right]^\prime &=& 0 \; , \label{bpaqseq} \\
     \left[ \beta_{\omega,j}^p(x) \beta_{\omega,j}^q(x) \right]^\prime &=& - 2C\omega I_j^2(x)  \; .  \label{bpbqseq}
\eea
\ees

Asymptotic boundary conditions at infinity can be obtained by solving Eqs.~\eqref{abprime} in the large $x$ limit.  Using the usual asymptotic expansions for Bessel functions,
we find solutions of the form
\bes\label{apbps}
\bea
     \alpha^p_{\omega,j}(x) &=&  \sum_{m = 0} a^p_m \, x^{C \omega -m}  \; ,  \\
     \beta^p_{\omega,j}(x) &=& e^{2 x} \sum_{m=0} b^p_m \, x^{C \omega -m -1} \; ,
\eea
\ees
and
\bes\label{aqbqs}
\bea
     \alpha^q_{\omega,j}(x) &=& e^{-2x} \sum_{m = 0}  a^q_m \, x^{-C \omega -m -1} \; , \\
     \beta^q_{\omega,j}(x) &=& \sum_{m=0} b^q_m \,x^{-C \omega -m} \;.
\eea
\ees
The $q$-solutions are chosen so that $F^q$ vanishes in the limit $x \rightarrow \infty$.  Using Eqs.~\eqref{apbps} and~\eqref{aqbqs} along with the constraint~\eqref{alphabetaconstraints},
we find the three boundary conditions:
\bes\label{bdys}
\bea
    \lim_{x \rightarrow \infty} \alpha^p_{\omega,j}(x) \alpha^q_{\omega,j}(x) &=&  0 \;, \label{bdysa} \\
    \lim_{x \rightarrow \infty} \alpha^p_{\omega,j}(x) \beta^q_{\omega,j}(x) &=& 1  \;, \label{bdysb} \\
    \lim_{x \rightarrow \infty} \alpha^q_{\omega,j}(x) \beta^p_{\omega,j}(x)  &=& 0 \;. \label{bdysc}
\eea
\ees
Note in particular that the boundary condition~\eqref{bdysb} implies that the solution to Eq.~\eqref{apbqseq} for all values of $x$ is
\be
\alpha^p_{\omega,j}(x) \beta^q_{\omega,j}(x) = 1 \;. \label{apbqssol}
\ee

The fourth boundary condition requires special care in this case.  It comes from the behavior of the $p$-solution, which must be regular at the inner boundary.  As in Sec.~V, we choose a distance $a$ which is large enough that the approximation~\eqref{metriclimit} can be assumed accurate for $r\stackrel{_>}{_\sim} a$, but which is small enough so that $a \ll r_e$, where $r_e$ is the value of $r$ at which we want to evaluate the stress-energy tensor.  We expect that at $r=a$ the contribution of $I_j$ and $K_j$ to $F_{\omega,j}^p$ will be comparable, so that
     \be \alpha_{\omega,j}^p(\omega a) I_j(\omega a) \sim \beta_{\omega,j}^p(\omega a) K_j(\omega a)  \; .\ee
Since $I_j(x)$ and $K_j(x)$ for small $x$ go as $x^j$ and $x^{-j}$ respectively, this tells us that
     \be \beta_{\omega,j}^p(\omega a)  \sim \left(\omega a\right)^{2j} \alpha_{\omega,j}^p(\omega a)\; . \ee
Multiplying both sides by $\beta_{\omega,j}^q(\omega a)$ and using Eq.~\eqref{apbqssol} gives
     \be\label{limits} \beta_{\omega,j}^p(\omega a) \beta^q_{\omega,j}(\omega a) \sim \left(\omega a\right)^{2j} \; . \ee
We anticipate that the largest contribution to the stress-energy tensor at $r=r_e$ will come from $\omega \sim r_e^{-1}$, and therefore expect that the right side of Eq.~\eqref{limits} will be proportional to $(a/r_e)^{2j}$.  For $j > \frc12$, this is negligible compared to $C\omega \sim C/r_e$ for large enough $r_e$, and hence we treat it as zero.   But for $j=\frc12$ this term could be comparable to (or larger than) $C\omega$.

For $j = \frc12$, it is useful to define the constant $\gamma$ such that
     \be\label{gamma-def} \gamma =  \lim_{\omega a \rightarrow 0} {\pi \beta_{\omega,1/2}^p(\omega a) \over \omega \alpha_{\omega,1/2}^p(\omega a)} \; .\ee
The factor of $\pi$ is included to simplify later relationships. The limit $a \omega \rightarrow 0$ is appropriate, because we expect $\omega \sim r_e^{-1}$, so this is equivalent to $r_e \gg a$.  Using Eq.~\eqref{apbqssol} we therefore find
\be
     \lim_{\omega a \rightarrow 0} \beta_{\omega,1/2}^p(\omega a) \beta_{\omega,1/2}^q(\omega a) = {\gamma \omega \over \pi} \lim_{\omega a \rightarrow 0} \alpha_{\omega,1/2}^p(\omega a) \beta_{\omega,1/2}^q(\omega a) = {\gamma \omega \over \pi} \; .
\ee
This yields our fourth boundary condition\footnote{It is conceivable that for certain pathological spacetime metrics this relationship will not hold, and hence the $\beta_j^p$ will be important for $j > \frc12$.  That is why our argument is not completely general.},
\be\label{bdysd}
     \lim_{x \rightarrow 0} \beta_{\omega,j}^p(x) \beta_{\omega,j}^q(x) = {\gamma \omega \over \pi} \delta_{j,1/2} \; .
\ee
It is important to note that $\gamma$ is of linear order in the metric perturbations $A$ and $B$.

Using the boundary conditions~\eqref{bdys} and \eqref{bdysd} one easily finds that the solutions to
 Eqs.~(\ref{scalarmodeproducts2}) are
\bes\label{scalarabproducts}
\bea
     \alpha_{\omega,j}^p(x)\alpha_{\omega,j}^q(x) &=& -2C\omega \int_x^\infty K_j^2(y) dy \; , \\
     \alpha_{\omega,j}^p(x)\beta_{\omega,j}^q(x) &=& 1 \; ,  \\
     \beta_{\omega,j}^p(x)\alpha_{\omega,j}^q(x) &=& 0 \; ,  \\
     \beta_{\omega,j}^p(x)\beta_{\omega,j}^q(x) &=& {\gamma\omega \over \pi} \delta_{j,1/2} -2C\omega \int_0^x I_j^2(y) dy \; .
\eea
\ees
Starting with Eq.~\eqref{F-def-inverse} and using Eqs.~\eqref{noprimedef} and~\eqref{scalarabproducts} gives
\bea\label{pq}
     p_{\omega\ell}(r)q_{\omega\ell}(r) &=& \left({1\over r} + {A \over r^2}\right) \biggl\{ I_j(x)K_j(x) - 2C\omega \left[K_j^2(x) \int_0^x I_j^2(y) dy  + I_j^2(x) \int_x^\infty K_j^2(y) dy \right]  \nn \\
     && \qquad {}  + {\gamma \omega \over 2x} e^{-2x} \delta_{j,1/2} \biggr\} - {B\omega \over r} \left[I_j(x)K_j^\prime(x) + K_j(x) I_j^\prime(x) \right] \; ,
\eea
where the explicit form for $K_{1/2}^2$, which is given in Eq.~(\ref{besselsmall}), has been used.

To compute $S_1$ and $S_5$, one can first multiply Eq.~(\ref{pq}) by $2\ell+1$, subtract $1/r + A/r^2$, and then sum over $\ell$ with the result that
\bea
     &&\sum_{\ell=0}^\infty \left[(2\ell{+}1) p_{\omega\ell}(r)q_{\omega\ell}(r) - {1 \over r} -{A \over r^2}\right] = \sum_{j=1/2} \biggl\{ - {4jC\omega \over r} \left[K_j^2(x)\int_0^x I_j^2(y) dy + I_j^2(x) \int_x^\infty K_j^2(y) dy \right]  \nn \\
     && \qquad {} + \left({1\over r} + {A \over r^2}\right) \left[ 2jI_j(x)K_j(x) - 1\right] - {2Bj\omega \over r} \left[I_j(x)K_j^\prime(x) + K_j(x) I_j^\prime(x) \right]  \biggr\} + {\gamma\omega\over 2rx} e^{-2x} \; .
\eea
Using the sum formulas~(\ref{sum1b}) and (\ref{sum1e}) along with the identity Eq.~(\ref{identA}) yields
\bea\label{s1s5sums}
     && \sum_{\ell=0}^\infty \left[(2\ell{+}1) p_{\omega\ell}(r)q_{\omega\ell}(r) - {1 \over r} - {A \over r^2} \right] \nn \\
     &&\qquad = - {C\omega \over r} \left(2x\int_x^\infty {dy \over y}e^{-2y} + 1 - e^{-2x} \right)- {x \over r} - {Ax \over r^2} + {B\omega \over r} + {\gamma \omega \over 2rx} e^{-2x} \nn  \\
     &&  \qquad = - {2C \over r^2} x^2\int_x^\infty {dy \over y}e^{-2y} +{C \over r^2} x e^{-2x} - {x \over r} - {2Ax \over r^2} + {\gamma \over 2r^2} e^{-2x} \; .
\eea
Then $S_1$ and $S_5$  can be found by substituting the above result into Eqs.~(\ref{S1def}) and (\ref{S5def}), and replacing $\omega$ with $x/r$:
\bes
\bea
     S_1 &=& {1 \over 4 \pi^2 r^5} \int_0^\infty dx \, x^2 \left( \frc12\gamma e^{-2x} - 2Cx^2\int_x^\infty {dy \over y}e^{-2y} +Cx e^{-2x} \right) \; , \\
     S_5 &=& {1 \over 4 \pi^2 r^3} \int_0^\infty dx \, \left( \frc12\gamma e^{-2x} - 2Cx^2\int_x^\infty {dy \over y}e^{-2y} +Cx e^{-2x} \right) \; .
\eea
\ees
The remaining integrals are straightforward to compute. The results are
\bes
\bea
     S_1 &=& {3C + 5 \gamma \over 160 \pi^2 r^5} \; , \label{S1val} \\
     S_5 &=& {C + 3 \gamma \over 48 \pi^2 r^3} \; . \label{S5val}
\eea
Next $S_4$ can be found from $S_5$ using the relation~(\ref{S4simple}) with the result that
\be
     S_4 = -{C + 3 \gamma \over 16 \pi^2 r^4} \; . \label{S4val}
\ee
\ees

The sum $S_3$ can be computed in a similar way by first substituting Eq.~(\ref{pq}) into Eq.~(\ref{S3def}) with the result that
\bea
     S_3 &=& \int_0^\infty {d\omega \over 4 \pi^2} \Biggl\{ \sum_{j=1/2}^\infty \Biggl[ \left({1\over r} + {A \over r^2}\right) \biggl\{ 2j^3I_j(x)K_j(x) - 4j^3 C\omega \biggl[K_j^2(x) \int_0^x I_j^2(y) dy  \nn \\
     && {} + I_j^2(x) \int_x^\infty K_j^2(y) dy \biggr] - j^2  \biggr\} - {2 B \omega j^3 \over r} \left[I_j(x) K_j^\prime(x) + I_j^\prime(x) K_j(x) \right]  + {\omega^2 r \over 2} + {3 A \omega^2 \over 2} \Biggr] \nn \\
     && {} + {\gamma \omega \over 8rx}e^{-2x} - {2 r^2 \omega^3 \over 3} - {8Ar \omega^3 \over 3} + {\omega \over 4} + {A \omega \over 6r} - {B \omega \over 3r} \Biggr\} \; .
\eea
Next replacing $\omega$ with $x/r$ and rearranging slightly one finds
\bea
     S_3 &=& {1 \over 4 \pi^2 r^3} \int_0^\infty dx \Biggl[  \sum_{j=1/2}^\infty \biggl\{(r+A) \left[ 2j^3I_j(x)K_j(x) - j^2 + \frc12x^2 \right]
      - Bx  \bigl[2j^3I_j(x) K_j^\prime(x)  \nn \\
     && {} + 2j^3I_j^\prime(x) K_j(x) +x \bigr]- Cx \left[ 4j^3K_j^2(x) \int_0^x I_j^2(y) dy  + 4j^3I_j^2(x) \int_x^\infty K_j^2(y) dy - x \right] \biggr\} \nn \\
     && {} + \frc18\gamma e^{-2x}  - \frc23rx^3 - \frc83Arx^3 + \frc14rx + \frc16Ax - \frc13Bx \Biggr] \; .
\eea\
Substituting Eqs.~(\ref{sum3a}), (\ref{sum3b}), and (\ref{identD}) for the remaining sums and integrals yields
\bea
     S_3 &=& {1 \over 4\pi^2r^3} \int_0^\infty dx \left\{ C \left[ \left( \frc43 x^4 - \frc12 x^2 \right) \int_x^\infty {dy\over y}e^{-2y} - \frc23x^3e^{-2x} + \frc13 x^2 e^{-2x} - \frc1{12} x e^{-2x} \right] + \frc18\gamma e^{-2x} \right\} \; . \nn \\ &&
\eea
Finally, computing the remaining integrals using integration by parts where necessary gives
\be\label{S3val}
     S_3 = {15 \gamma - 7C \over 960 \pi^2 r^3} \; .
\ee

To find $S_2$, it is easiest to use Eq.~\eqref{S2simple}.  To leading order this relationship is
\be
     S_2 = {1 \over 2} {dS_4 \over dr} + {S_4 \over r} - S_1 - {S_3 \over r^2} + {S_5 \over 4r^2} \; .
\ee
Substituting explicit forms from Eqs.~(\ref{S1val}), (\ref{S5val}), (\ref{S4val}), and (\ref{S3val}), we find
\be\label{S2val} S_2 = {9C+25\gamma  \over 160 \pi^2 r^5} \; . \ee
It is now straightforward to substitute all of our expressions for $S_n$ into Eq.~(\ref{TabSns}) to obtain the full expression for $\langle {T^a}_b \rangle$:
\be\label{Tmunuboson}
     \langle {T^a}_b \rangle = {1 \over 80 \pi^2 r^5}\left\{ C {\rm diag}\left(-1,2,-3,-3\right) + 5 \left[\xi C + 3 (\xi - \frc16) \gamma  \right] {\rm diag}
     \left(2,-2,3,3\right) \right\}\; .
\ee
In general this result depends on the details of the geometry at small $r$ through the parameter $\gamma$, but for a conformally invariant scalar field ($\xi=\frc16$) the $\gamma$-dependance cancels.  We can also compute the quantity $\langle \phi^2 \rangle$ by using Eqs.~\eqref{phi2num} and \eqref{S5val}, keeping in mind that the numerical contribution dominates the analytical contribution at large $r$, so that
     \be\label{phi2val} \langle \phi^2 \rangle = S_5 = {C + 3 \gamma \over 48 \pi^2 r^3} \; . \ee
Note that unlike the stress-energy tensor, the value of $\langle \phi^2 \rangle$ depends on $\gamma$ for any value of $\xi$.

To evaluate Eq.~\eqref{Tmunuboson} or \eqref{phi2val} explicitly, we need to determine $\gamma$.  To do so it is necessary to solve, analytically or numerically, the exact mode equation~(\ref{mode-eq-exact})
when $\omega = \ell = 0$ for the $p$-function.  This equation is
\be\label{p00mode}
     {d^2p_{00} \over dr^2} + \left({2\over r} + {1 \over 2f} {df \over dr} - {1 \over 2h} {dh \over dr} \right) {dp_{00} \over dr} - \xi R h p_{00} = 0 \; .
\ee
In the asymptotic region, the metric becomes flat, and the solution takes the form
\be\label{p00limit}
     p_{00}(r) = c_0 + {c_1/r} \;.
\ee
Substitution into Eq.~(\ref{F-def}) then gives the function $F^p$, which we match to order $1/r$ with the general form Eq.~(\ref{noprimedef}):
\bea
    F_{00}^p &=& r^{1/2} c_0 \left(1 + {c_1 \over c_0 r} + {B-A \over 2r}  \right) \nn \\
             &=& \lim_{\omega \rightarrow 0} \left[\alpha_{\omega,1/2}^p(\omega r) I_{1/2}(\omega r) + \beta_{\omega,1/2}^p(\omega r) K_{1/2}(\omega r) \right] \nn \\
             &=& r^{1/2} \left\{ \lim_{\omega \rightarrow 0} \left[ \alpha_{\omega,1/2}^p(\omega r) \left(\frac{2 \omega}{\pi} \right)^{1/2} \right]  \right\} \left( 1 + \frac{\gamma}{2 r} \right)
  \;.  \label{F00p}
\eea
Here the definition~\eqref{gamma-def} has been used.  Clearly
\begin{subequations}
\bea
&& \lim_{\omega \rightarrow 0} \left[ \alpha_{\omega,1/2}^p(\omega r) \left(\frac{2 \omega}{\pi} \right)^{1/2} \right] = c_0   \;, \\
  &&  \gamma = {2c_1 \over c_0} + B-A \; .\label{gammasolve}
\eea
\end{subequations}
Thus $\gamma$ can be found by solving Eq.~(\ref{p00mode}) either numerically or analytically and matching the solution with its asymptotic form~(\ref{p00limit})
to obtain the values of $c_0$ and $c_1$.

For example, consider the case of a Schwarzschild black hole, with metric functions given by Eqs.~\eqref{Schwarz}, so that $A = B = M$.  Since the scalar curvature $R$ is zero everywhere, there are two linearly independent solutions to Eq.~(\ref{p00mode}), one of which diverges at the  horizon, and the other of which is constant everywhere.  The latter is the $p$-solution, and we see from Eqs.~(\ref{p00limit}) and (\ref{gammasolve}) that $\gamma = 0$.  The resulting stress-energy tensor is given in \eqref{Tmnsch} for all values of $\xi$.  Thus our use of the WKB approximation in Section~\ref{sec:WKB} gives the correct leading
order behavior for the stress-energy tensor in the far field region.

In contrast, consider a nonrelativistic star, by which we mean a static spherically symmetric weak field source whose stress-energy tensor is dominated by ${T^t}_t = -\rho(r)$.  Such a source will have $R \approx 8\pi \rho$. In flat space $p_{00}$ is equal to a constant.  In the weak field limit we can therefore write
\be p_{00} = c_2 + \delta p \ee
and substitute it into the mode equation~(\ref{p00mode}).  Note that without loss of generality we can set $\delta p=0$ at the origin.  Assuming that $\delta p$ is small and keeping only first order terms gives
\be
     {d \over dr}\left( r^2 {d \delta p \over dr} \right) = 8\pi \xi r^2 \rho(r) c_2  \; .
\ee
For points outside the star this can be integrated twice to obtain
\be
      \delta p = c_3  - {2 \xi M c_2 \over r} \;,
\ee
where $M$ is the total mass and  $|c_3| \ll |c_2|$.  Matching to the form~(\ref{p00limit}), we see that $c_0 = c_2+c_3$.
Then to leading order, $c_1 = -2\xi M c_0$.  Using Eq.~(\ref{gammasolve}) with $A = B = M$ gives $\gamma = -4 \xi M$.  Thus to leading order
\begin{subequations}
\bea
     \langle \phi^2 \rangle_{\rm Sch} &=& {M \over 24 \pi^2 r^3} \;, \\
      \langle \phi^2 \rangle_{\rm nr}  &=& {M \over 24 \pi^2 r^3} (1 - 6 \xi) \;, \\
     \langle {T^a}_b \rangle_{\rm Sch} &=& {M \over 40 \pi^2 r^5}\left\{ {\rm diag}\left(-1,2,-3,-3\right) + 5 \xi {\rm diag} \left(2,-2,3,3\right) \right\}\; , \label{Tabscalarsch} \\
     \langle {T^a}_b \rangle_{\rm nr} &=& {M \over 40 \pi^2 r^5}\left\{ {\rm diag}\left(-1,2,-3,-3\right) + \left[5 \xi - 30 \xi \left(\xi-\frc16 \right) \right] {\rm diag} \left(2,-2,3,3\right) \right\} \; .
\eea
\end{subequations}
These are the results found previously in Refs.~\cite{mazz} and~\cite{abf}.  As shown in~\cite{af}, to leading order at large $r$ the quantity $\langle \phi^2 \rangle$ in Schwarzschild spacetime differs from its value outside of a static spherically symmetric star
except in the case  $\xi = 0$, while $\langle {T^a}_b \rangle$ differs from its value outside of a static spherically symmetric star except in the cases $\xi = \frc16$ and $\xi=0$.

The argument that for $\xi=0$ the asymptotic values of $\langle \phi^2 \rangle$ and $\langle {T^a}_b \rangle$ are the same outside a nonrelativistic star as for a Schwarzschild black hole can be generalized to the statement that for $\xi = 0$ the asymptotic values of these quantities depend only on the leading order geometry in the far field region for any star or black hole.  The exact solution is in this case simply
\begin{subequations}
\bea
  p_{00}(r) &=& c \; ,  \label{pconst} \\
  q_{00}(r) &=& {1 \over c} \int_r^\infty {dr^\prime \over r^{\prime 2}} \sqrt{h(r^\prime) \over f(r^\prime)} \; .
\eea
\end{subequations}
Here it is clear that $p_{00}$ must be the constant solution because it is the only solution that does not diverge at $r = 0$ in the case of a star, or at the event horizon in the case of a black hole. Then matching to Eq.~\eqref{p00limit} gives $c_1 = 0$, and hence $\gamma = B - A$.

The above result, while quite general, does not hold for a wormhole.  By a wormhole we mean a metric such that at some throat radius $b$, $h(r)$ has a simple pole while $f(r)$ is finite and well-behaved.  This represents a connection between two asymptotically flat universes, which for simplicity we will assume have identical metric functions $h(r)$ and $f(r)$.  The radius $r=b$ is the minimum radius for which the metric is defined and represents the point at which the two universes join.  For minimal coupling, $\xi =0$, the $p$ solution will be the one that vanishes at infinity in the other universe, and therefore it is not the constant solution of Eq.~\eqref{pconst}, but instead the solution that behaves like $r^{-1}$ at large $r$ in that universe.  Note that unlike the
case when there is an event horizon, $\int \sqrt{h/f} dr$ is finite at the throat of the wormhole.
The exact solution to Eq.~\eqref{p00mode} will satisfy
\be\label{p00soln1}
   {d p_{00} \over dr} = \pm {c \over r^2} \sqrt{h(r) \over f(r) } \; ,
\ee
where $c$ is an unknown constant.  Care must be taken at the coordinate singularity $r=b$.  A temporary change of coordinates to $r=b+s^2$ shows that the singularity in Eq.~\eqref{p00soln1} is integrable, so that $p_{00}$ is continuous, but $dp_{00}/dr$ changes sign as you cross the throat. In Eq.~\eqref{p00soln1}, the $\pm$ refers to which universe you are in, with the $+$ sign corresponding to our universe and the $-$ sign to the other universe.  Keeping in mind that $p_{00}(\infty) = 0$ in the other universe, one can integrate from $\infty$ in that universe to
 the throat $b$ of the wormhole and then in our universe integrate out from the the throat to a radius $r_e$ with the result that
\be
    p_{00}(r_e) = -c \int_\infty^b {dr \over r^{2}} \sqrt{h(r) \over f(r) } + c \int_b^{r_e} {dr \over r^{2}} \sqrt{h(r) \over f(r) } = 2c\int_b^\infty {dr \over r^{2}} \sqrt{h(r) \over f(r) } - c \int_{r_e}^\infty {dr \over r^{2}} \sqrt{h(r) \over f(r)} \; .
\ee
If $r_e$ is chosen so that it is large enough that we can treat the metric as flat for $r > r_e$ then
\be
   p_{00}(r_e) = 2c\int_b^\infty {dr \over r^{2}} \sqrt{h(r) \over f(r) } - {c \over r_e} \; .
\ee
Comparison with Eqs.~\eqref{p00limit} and \eqref{gammasolve} shows that, for a scalar field with minimal curvature coupling $\xi=0$,
\be
   \gamma = - \left[\int_b^\infty {dr \over r^2} \sqrt{h(r) \over f(r)} \right]^{-1} + B - A \; .
\ee

\section{Numerical Computations}
\label{sec:numeric}

There are several cases in which we have done numerical computations of the stress-energy tensor for massless fields far from the event horizon
of a black hole or the throat of a wormhole.  These are important because they are computations of the full renormalized stress-energy tensor
for the quantum fields and thus serve as important checks on the WKB calculation and the exact analytic calculations in the previous sections.

The first numerical computation of the stress-energy tensor for a scalar field in the Boulware state was done in the case of conformal coupling in~\cite{jmo}.  However, it is not possible to deduce the large $r$ behavior of the stress-energy tensor from the plots in that paper.  In~\cite{abf} some results of numerical calculations for scalar fields with arbitrary coupling $\xi$ to the scalar curvature were shown.  As discussed in that paper, the results agree with those in Eq.~\eqref{Tabscalarsch} to within at least two digit accuracy.

\begin{figure}
\centering
\subfloat{\includegraphics[angle=90,width=3.1in,clip]{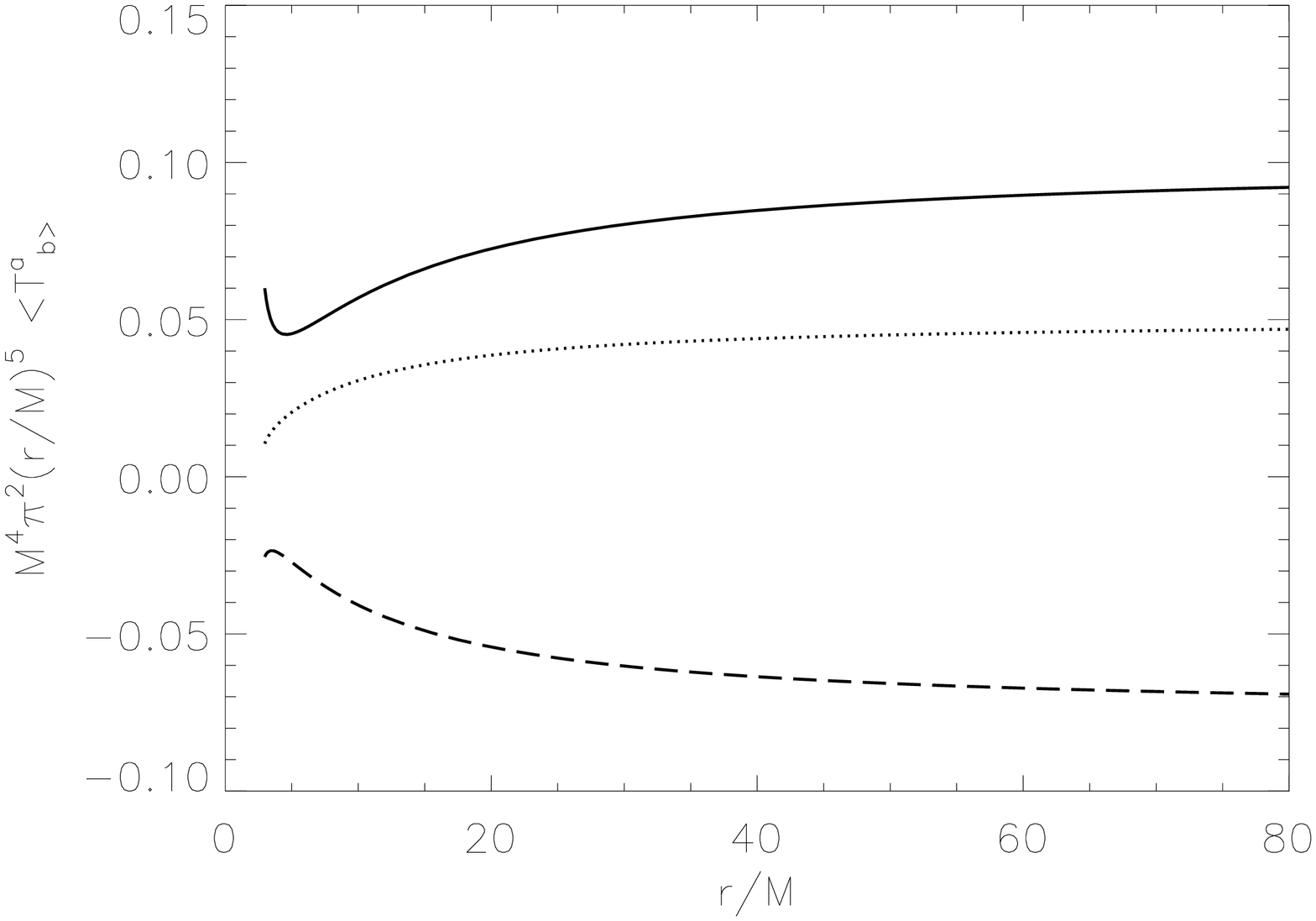}}\hfill
\subfloat{\includegraphics[angle=90,width=3.1in,clip]{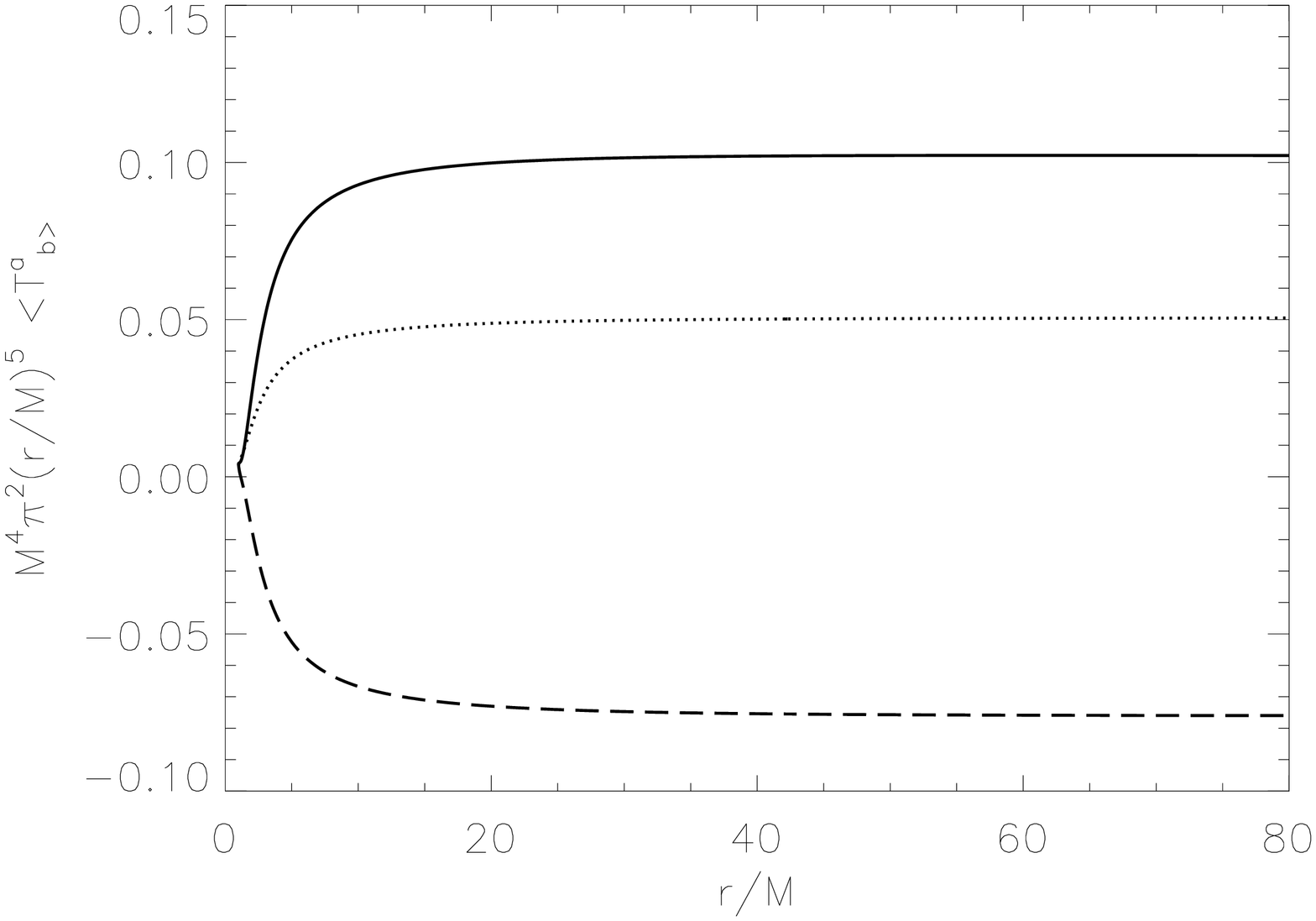}}
\caption{Numerical calculation of the stress-energy tensor for the massless spin $1/2$ field in Schwarzschild (left) spacetime and the \ern\ (right) spacetime
with the renormalization parameter $\mu = M^{-1}$ in the notation of Ref.~\cite{gac}.
In each plot, the solid curve is $\langle {T^t}_t \rangle$, the dotted curve is $\langle {T^r}_r \rangle$,
and the dashed curve is $\langle {T^\theta}_\theta \rangle = \langle {T^\phi}_\phi \rangle$.}
\end{figure}

For the massless spin $1/2$ field we have computed $\langle {T^a}_b \rangle$ in Schwarzschild and extreme Reissner-Nordstr\"om spacetimes.  It is clear from the plots in Fig.\ 1 that the large $r$ behavior
of the components goes like $M/r^5$.  Since the metric for the extreme Reissner-Nordstr\"om spacetime is
\be  f = h^{-1} = 1 - \frac{2 M}{r} + \frac{M^2}{r^2}  \ee
the values of $A$ and $B$ in Eq.~\eqref{metriclimit} are $A = B = M$,  the same as for Schwarzschild spacetime.  Thus the leading order behavior of the stress-energy tensor predicted
by Eq.~\eqref{Tmunufermi3} for both Schwarzschild and Reissner-Nordstr\"om spacetimes is
\be\label{Tschern} \langle {T^a}_b \rangle = {M \over40\pi^2r^5} \, {\rm diag}\left(4,2,-3,-3\right) \; . \ee
Fitting the numerical data to a series in inverse powers of $r$, we find that in each case there is agreement with the results in
Eq.~\eqref{Tschern} to at least two digits.

Numerical computations have also been done for three different wormhole metrics.  In each case, the throat is at $r=b$.  The explicit form of the metric functions $f$ and $h^{-1}$, together with the corresponding coefficients $A$ and $B$ and the predictions of Eq.~\eqref{Tmunufermi3} are
given in Table 1.  The stress-energy tensors for these metrics are shown in Fig.\ 2.  It is clear from the figures that the large $r$ behavior of the stress-energy again goes like $1/r^5$.  Fitting the numerical data to a power series in inverse powers of $r$ we find in each case agreement with the analytic results displayed in Table 1 to within approximately two digits.

\begin{figure}
\centering
\subfloat{\includegraphics[angle=90,width=3.1in,clip]{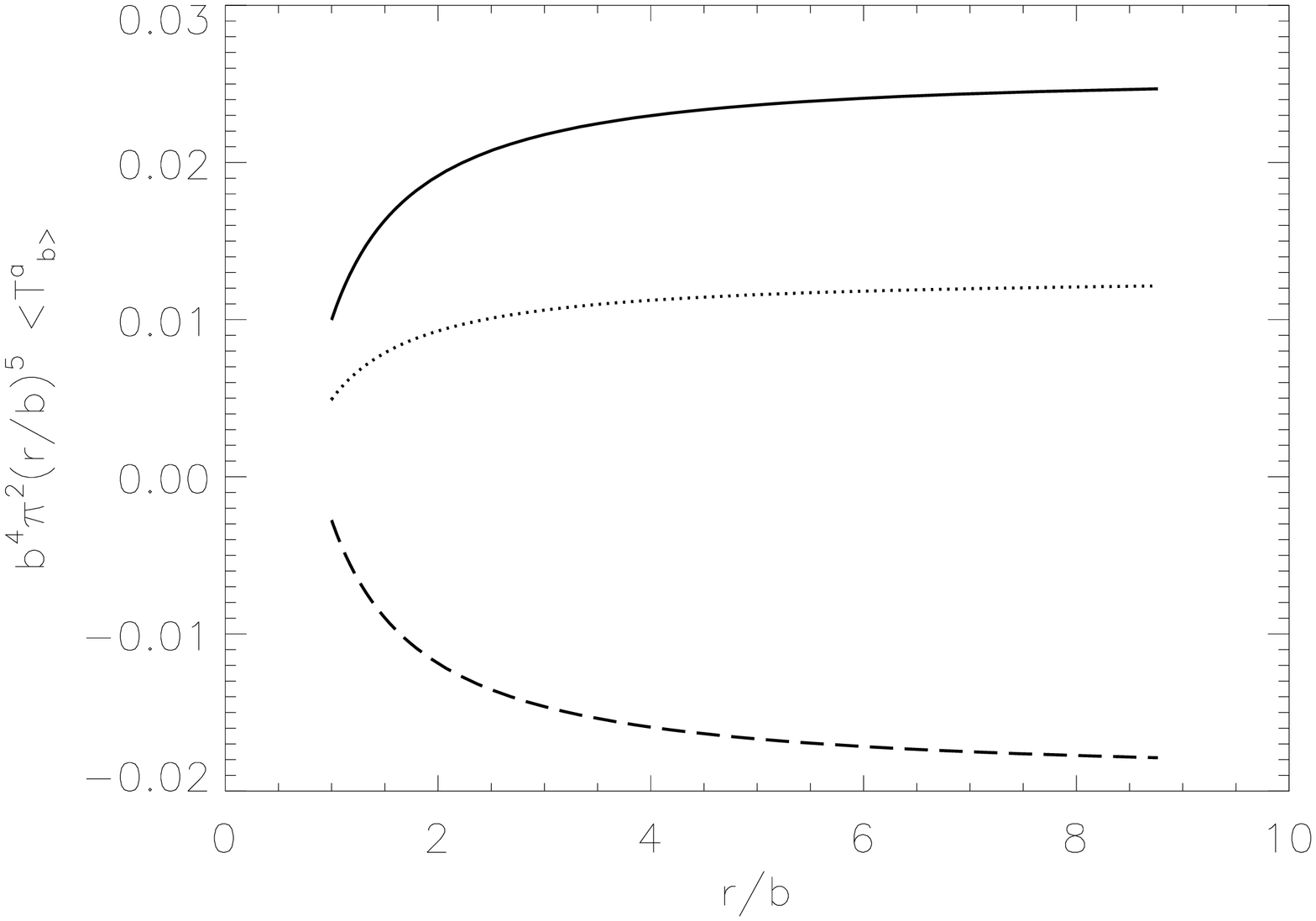}}\hfill
\subfloat{\includegraphics[angle=90,width=3.1in,clip]{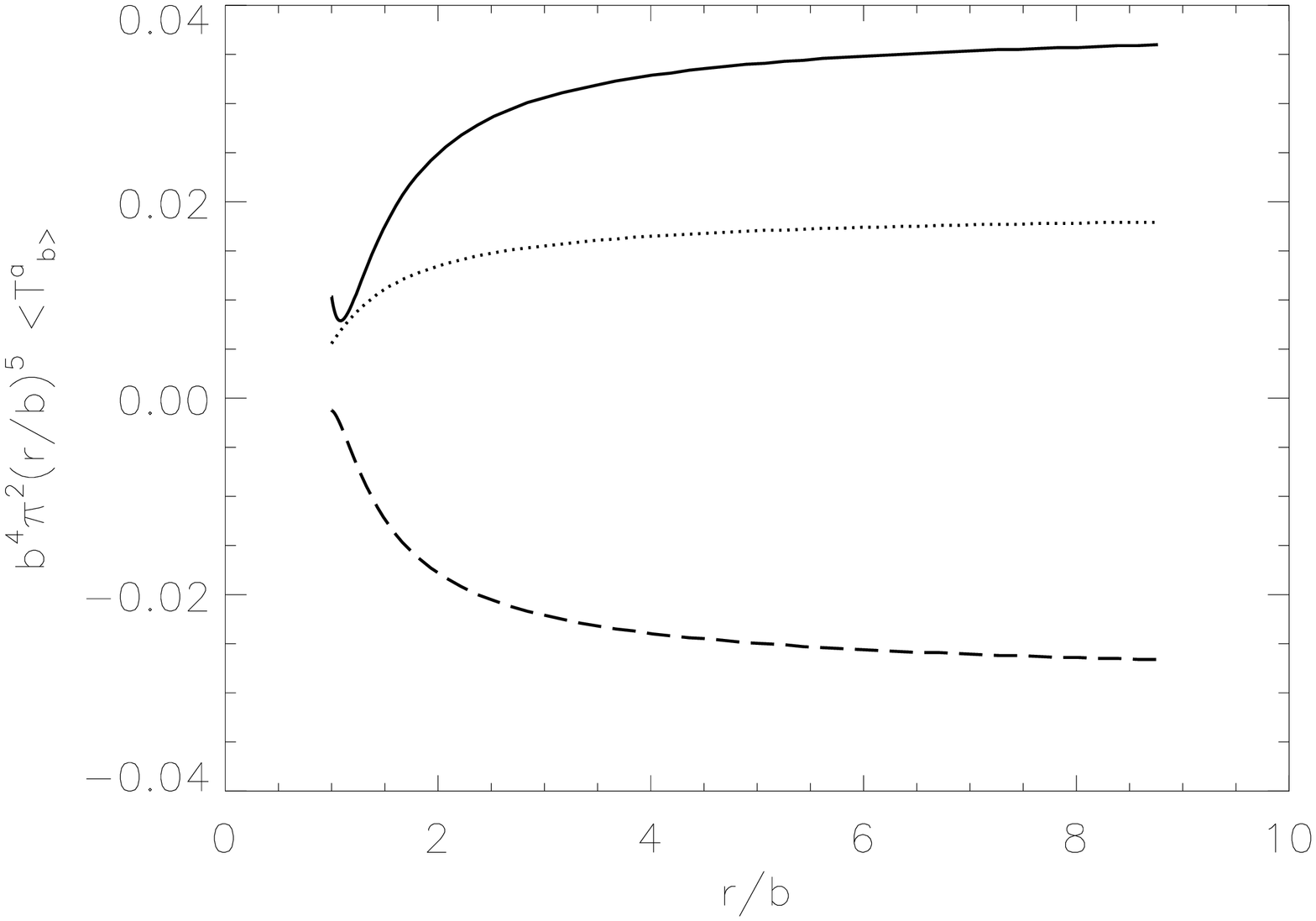}}\hfill
\subfloat{\includegraphics[angle=90,width=3.1in,clip]{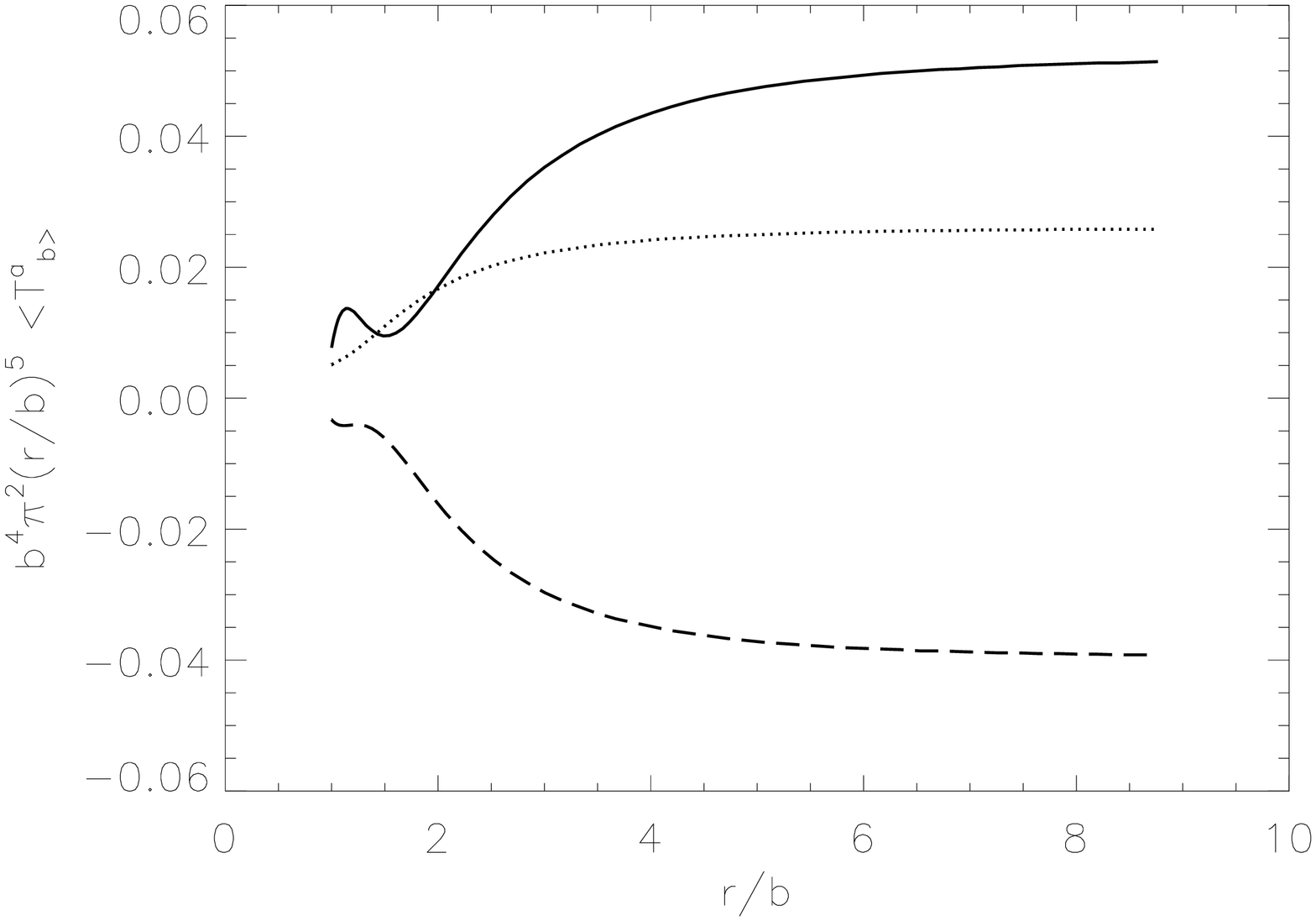}}
\caption{Numerical calculation of the stress-energy tensor for the massless spin $1/2$ field in three wormhole metrics with renormalization parameter $\mu = b^{-1}$ in the notation of Ref.~\cite{gac}.  All three have $h(r)^{-1} = 1-b/r$.  The metric function $f(r)$ is given by $f(r) = 1$ (top left), $f(r)=1-b/2r$ (top right) and $f(r) = 1-b/r+b/2r^2$ (bottom).   In each plot the solid curve is $\langle {T^t}_t \rangle$, the dotted curve is $\langle {T^r}_r \rangle$, and the dashed curve is $\langle {T^\theta}_\theta \rangle = \langle {T^\phi}_\phi \rangle$.}
\end{figure}

{\center{
\begin{table}
 \begin{tabular}{|c|c|c|c|c|}
      \hline
      ~$f$~            & ~$h^{-1}$~ & ~$A$~   & ~$B$~   & ~$320\pi^2r^5\langle {T^a}_b\rangle$~\\ \hline
      ~$1$~            & ~$1-b/r$~  & ~$0$~   & ~$b/2$~ & ~$2b \,{\rm diag}(4,2,-3,-3)$~  \\ \hline
      ~$1-b/2r$~       & ~$1-b/r$~  & ~$b/4$~ & ~$b/2$~ & ~$3b \,{\rm diag}(4,2,-3,-3)$~ \\ \hline
      ~$1-b/r+b^2/2r^2$~ & ~$1-b/r$~  & ~$b/2$~ & ~$b/2$~ & ~$4b \,{\rm diag}(4,2,-3,-3)$~ \\ \hline
  \end{tabular}
  \caption{Leading order stress-energy tensors for three wormhole metrics.}
  \label{restab}
\end{table}
}}

\section{Comparison with Previous Calculations for Schwarzschild Spacetime}
\label{sec:numsch}

It is possible to compare our results for both the scalar and spin $1/2$ fields with previous computations of the leading order corrections to the gravitational potential $\Phi$ in Schwarzschild
spacetime which have been found by computing one loop corrections to the graviton propagator.  Such corrections have been computed by Duff~\cite{duff} for both the
massless conformally coupled scalar field and the massless spin $1/2$ field and by Hamber and Liu~\cite{h-l} for the massless minimally coupled scalar field.

First note that the one loop correction to the graviton propagator for a given type of quantum field gives the
leading order correction to the gravitational potential $\Phi$ from that field~\cite{duff}.  Here $\Phi$ is defined by the relation
\be  g_{tt} = -(1 - 2 \Phi) \;. \label{Phi-def} \ee
Clearly at the classical order $\Phi = M/r$.

Another way to obtain quantum corrections to $\Phi$ is to solve the linearized semiclassical
backreaction equations.  This has been done previously for massless scalar fields with arbitrary coupling to the scalar curvature in Schwarzschild
spacetime~\cite{abf} and in the exterior region of a nonrelativistic static spherically symmetric star~\cite{mazz}.
If the metric functions in Eq.~\eqref{metric} are written as
\begin{subequations}
\bea
f(r) &=& \left[1-\frac{2m(r)}{r}\right]e^{2\phi(r)}  \;, \\
h(r) &=& \left[1-\frac{2m(r)}{r}\right]^{-1} \;,
\eea
\end{subequations}
then the semiclassical backreaction equations take the simple form
\begin{subequations}
\bea \label{backeq1}
 \frac{d m(r)}{dr} &=& -4\pi r^2 \langle {T^t}_t \rangle  \ ,  \\
\label{backeq2}
 \frac{d \phi(r)}{dr} &=& 4\pi r \frac{ \langle {T^r}_r\rangle - \langle {T^t}_t\rangle
 }{1-2m(r)/r}\; .
 \eea
\label{backeq}
\end{subequations}

Substituting Eq.\ \eqref{Tabscalarsch} into Eqs.\ \eqref{backeq}  and
integrating gives
\begin{subequations}
\begin{eqnarray}
m(r) &=& M \left[1 - \frac{(1-10\xi)}{20 \pi r^2} \right] + O\left(\frac{1}{r^3} \right) \;,  \\
\phi(r) &=& -\frac{M (3 - 20 \xi)}{30 \pi r^3}  + O\left(\frac{1}{r^4} \right) \;,
\end{eqnarray}
\end{subequations}
which in turn implies that
\be \Phi = \frac{M}{r} + \left[\frac{1}{45\pi} - \left(\xi-\frac{1}{6}\right)\frac{1}{6\pi} \right] \frac{M}{r^3} + O\left(\frac{1}{r^4}\right) \;. \ee
For $\xi = 1/6$ this gives the same value as that found by Duff~\cite{duff} and for $\xi = 0$ it gives the value found by Hamber and Liu~\cite{h-l}.
The same agreement was found previously in Ref.~\cite{abf} using our numerical results and the WKB calculation discussed in Section \ref{sec:WKB}.

For the massless spin $1/2$ field, substituting Eq.~\eqref{Tschern} into  Eqs.~\eqref{backeq}  and
integrating gives
\begin{subequations}
\begin{eqnarray}
m(r) &=& M \left(1 + \frac{1}{5 \pi r^2} \right) + O\left(\frac{1}{r^3} \right) \;,  \\
\phi(r) &=& \frac{M}{15 \pi r^3}  + O\left(\frac{1}{r^4} \right) \;,
\end{eqnarray}
\end{subequations}
and
\be \Phi = \frac{M}{r} + \frac{2M}{15\pi r^3} + O\left(\frac{1}{r^4}\right) \;. \ee
This value for the correction to $\Phi$ is twice as large as that found by Duff~\cite{duff} for the spin $\frc12$ field.  This is due to the fact that Duff considered a two component field and we are considering a four component one.

One would expect the graviton propagator computations discussed above to give the correct changes in the gravitational potential in the weak field approximation, {\it i.e.} for the exterior of a static star.  The reason that we get the same corrections for massless fields in Schwarzschild spacetime is that for the conformally coupled scalar field and the spin $1/2$ field
the leading order stress-energy depends only on the local geometry in the far field limit.  As discussed previously, for the minimally coupled scalar field the leading order stress-energy at large $r$ is the same for a static star as for a Schwarzschild black hole.

\section{Discussion}
\label{sec:conclusions}

In this paper analytic expressions have been computed for the leading order asymptotic behaviors of the quantities $\langle \phi^2 \rangle$ and $\langle {T^a}_b \rangle$ for arbitrarily coupled massless spin $0$ fields, and of $\langle {T^a}_b \rangle$ for the massless spin $\frc12$ field, in the case that the fields are in the Boulware state.  The results are expected to be valid for most static spherically symmetric spacetimes for which the metric has the asymptotic form~\eqref{metriclimit}.  As discussed in detail in Sections~\ref{sec:Fermions} and~\ref{sec:Bosons}, there may be some spacetimes, which are probably pathological, for which our results are not valid.  For this reason it is important to compare these analytic results with those obtained in other completely independent calculations.

For massless scalar fields with arbitrary coupling $\xi$ to the scalar curvature in Schwarzschild spacetime a combination of the WKB approximation for the modes and conservation of the stress-energy tensor has been used in Section III to compute the asymptotic behaviors of the quantities $\langle \phi^2 \rangle$ and $\langle {T^a}_b \rangle$.  The results are in agreement with the more general analytic calculations mentioned above.

The asymptotic behaviors of  $\langle \phi^2 \rangle$ and $\langle {T^a}_b \rangle$ in the region outside of a nonrelativistic, static spherically symmetric star have been computed analytically using different methods than ours in Ref.~\cite{mazz}.  Their results are in complete agreement with ours.

The asymptotic behavior of the gravitational potential as defined in Eq.~\eqref{Phi-def}
has been obtained by computing one loop corrections to the graviton propagator for massless scalar fields with conformal~\cite{duff} and minimal coupling~\cite{h-l} as well as the massless
spin $\frc12$ field~\cite{duff}.  Another way to obtain the corrections to the gravitational potential is to solve the linearized semiclassical backreaction equations using the analytic results for the asymptotic behavior of the stress-energy tensor.
This has been done previously for massless scalar fields with arbitrary coupling to the scalar curvature in Schwarzschild
spacetime~\cite{abf} and in the exterior region of a nonrelativistic static spherically symmetric star~\cite{mazz}.  In both cases the results agree with the one loop calculations in~\cite{duff,h-l}.
This is to be expected since there is no difference in the leading order behavior of the stress-energy tensor in Schwarzschild spacetime versus that of the exterior region of a nonrelativistic static spherically symmetric star for the cases of conformal or minimal coupling~\cite{af}.  In Section VIII we have solved the linearized semiclassical backreaction equations for the massless spin $\frc12$ field and found agreement with the corresponding one loop calculation in~\cite{duff}.

Numerical computations of both $\langle \phi^2 \rangle$ and $\langle {T^a}_b \rangle$ have been made for massless scalar fields with arbitrary coupling to the scalar curvature in Schwarzschild spacetime~\cite{abf}. In these calculations the full stress-energy tensor has been computed.  This has also been done here for the massless spin $\frc12$ field in Schwarzschild spacetime, the extreme Reissner-Nordstr\"{o}m spacetime, and three wormhole spacetimes.  The results are displayed and discussed in Section VII.
By fitting the numerical data to power series in $1/r$ we have found agreement with the analytic results discussed above.

As mentioned in the introduction, the stress-energy tensor for quantized fields is intrinsically nonlocal.
An interesting question that was raised in the comparison between static stars and Schwarzschild black holes in Ref.~\cite{af} is whether the leading
order behavior of the stress-energy tensor in the asymptotic region of a static, spherically symmetric, asymptotically flat spacetime is local or nonlocal.  As stated in the introduction, by
local we mean that the leading order terms in the stress-energy tensor depend only on the leading order deviations of the metric from a flat space metric.
In Ref.~\cite{af} it was found that when the geometry at large values of $r$ is the Schwarzschild geometry, the
leading order behavior of the stress-energy tensor for massless scalar fields is nonlocal except in the special cases of conformal and minimal coupling.  This result was extended in Ref.~\cite{mazz2} to the $D \ge 4$ dimensional Schwarzschild-Tangherlini geometry.

Examination of Eqs.~\eqref{Tmunufermi3} and~\eqref{Tmunuboson}  indicate that in the asymptotic region the stress-energy tensor is always local for the massless spin $\frc12$ field and for the conformally coupled massless scalar field.  We believe that these results combine with those of Ref.~\cite{mazz2} to give strong evidence that the locality of the asymptotic behavior of the stress-energy tensor is a consequence of the conformal symmetry exhibited by these fields.
For the case of the minimally coupled massless scalar field the leading order asymptotic behavior of the stress-energy tensor is local for any static spherically symmetric spacetime which contains either a star or a black hole, but it is nonlocal if the spacetime contains a wormhole.
Thus in many cases of physical interest the leading order behavior of the stress-energy tensor in the asymptotically flat region depends only on the local geometry there.

\section*{Acknowledgments}
P.R.A. would like to thank R. Wald and L. Ford for helpful comments.  P.R.A., A.F., and S.F. would like to thank R. Balbinot for helpful discussions.
The work of P.R.A. was supported in part by the National Science Foundation under
Grant Nos.\ PHY-0556292 and PHY-0856050.  A.F. acknowledges financial support by Generalitat Valenciana and MICINN
Grant No.\ FIS2008-06078-C03-02.  S.F.'s research is supported by the Anne McLaren
fellowship.  Numerical computations were performed on the Wake Forest University DEAC Cluster with
support from an IBM SUR grant and the Wake Forest University IS Department. Computational
results were supported by storage hardware awarded to Wake Forest University through
an IBM SUR grant.

\appendix*
\section{Bessel Function Identities}
\label{bessel}

In Sections \ref{sec:Fermions} and \ref{sec:Bosons} expressions for the stress-energy tensors of massless spin $1/2$ and spin $0$ fields are derived
which contain sums of products of the modified Bessel functions $I_j(x)$ and $K_j(x)$, with half-integer values for the index $j$.
In this Appendix we derive identities which make it possible to compute those sums.  Throughout it is assumed that $j$ takes on only half-integer
 values.  The functions $I_j(x)$ and $K_j(x)$ are linearly independent solutions
to the equation
\be\label{besseldef} f^{\prime\prime}(x)+{1\over x}f^\prime(x)-\left(1+{j^2\over x^2}\right)f(x) = 0 \; . \ee
For $j = \pm \frc12$ they are
\be\label{besselsmall}
    I_{-1/2}(x) = {2 \cosh x \over \sqrt{2 \pi x}} \; , \qquad I_{1/2}(x) = {2 \sinh x \over \sqrt{2 \pi x}} \; , \qquad K_{-1/2}(x) = K_{1/2}(x) = \sqrt{\pi \over 2x} e^{-x} \; .
\ee
All other half-integer values can be computed using the recursion relations
\bea\label{recursion}
    I_{j \pm 1}(x) &=& I'_j(x) \mp {j \over x}I_j(x) \; , \nn \\
    K_{j \pm 1}(x) &=& - K'_j(x) \pm {j \over x} K_j(x) \; .
\eea
These functions also satisfy the Wronskian condition
\be\label{Wronsk1} I_j(x)K'_j(x)-K_j(x)I'_j(x)=-{1 \over x} \; , \ee
which, with the help of the recursion relations, can be rewritten as
\be\label{Wronsk2} x\left[I_j(x)K_{j+1}(x)+K_j(x)I_{j+1}(x)\right] = 1 \; . \ee
 For $0<x \ll \sqrt{j}$ and $j > 1$,
\bea\label{smallx}
    I_j(x) &=& {1 \over \Gamma(j{+}1)} \left(x\over2\right)^j \left[ 1 + O\left(\frac{x^2}{j} \right) \right] \; , \nn\\
    K_j(x) &=& {\Gamma(j)\over2} \left(2\over x\right)^j \left[ 1 + O\left(\frac{x^2}{j} \right) \right]  \; ,
\eea
while for $x \gg j^2$,
\bea\label{bigx}
    I_j(x) &=& {1 \over \sqrt{2\pi x}}e^x \left[1 + O\left(j^2 \over x \right) \right]   \; , \nn \\
    K_j(x) &=& \sqrt{\pi \over 2x} e^{-x}  \left[1 + O\left(j^2 \over x \right) \right]          \; .
\eea

One of the sums which we need is
\be Q_1 \equiv \lim_{J \rightarrow \infty} \sum_{j=1/2}^J \left[I_j(x) K_{j-1}(x) + I_{j+1}(x) K_j(x) \right] \; .  \ee
To evaluate this sum first multiply by $x$, then take the derivative with respect to $x$ and use the recursion relations~\eqref{recursion} to obtain
\be {d \over dx} (xQ_1) = \lim_{J \rightarrow \infty} \sum_{j=1/2}^J x\left[ I_{j-1}(x)K_{j-1}(x)-I_{j+1}(x)K_{j+1}(x) \right] \; . \ee
Most of the terms cancel, as can be seen by shifting the indices on the two sums.  The result is
\bea
    {d \over dx} (xQ_1) &=& \lim_{J \rightarrow \infty} \left[\sum_{j=-1/2}^{J-1} xI_j(x)K_j(x) - \sum_{j=3/2}^{J+1} xI_j(x)K_j(x) \right] \nn \\
    &=& xI_{-1/2}(x)K_{-1/2}(x) + xI_{1/2}(x)K_{1/2}(x) - \lim_{J \rightarrow \infty}\left[xI_{J+1}(x)K_{J+1}(x) + xI_J(x)K_J(x) \right] \; .\nn \\
    &&
\eea
Using Eqs.~\eqref{besselsmall} and~\eqref{smallx} one finds that
\be\label{sumderiv1} {d \over dx} (xQ_1) = 1 \; .  \ee
Integrating this equation from 0 to $x$, and dividing the result by $x$ yields the identity
\be\label{sum1a} Q_1 = \sum_{j=1/2}^\infty \left[I_j(x) K_{j-1}(x) + I_{j+1}(x) K_j(x) \right] = 1 \; . \ee
Using the recursion relations~(\ref{recursion}) and the Wronskian condition~(\ref{Wronsk1}) then gives the closely related identity
\be\label{sum1b}  \sum_{j=1/2}^\infty [2jI_j(x)K_j(x)-1] = -x  \;. \ee
A third identity can be found by taking the derivative to obtain
\be\label{sum1e} \sum_{j=1/2}^\infty 2j[I_j^\prime(x)K_j(x) + I_j(x)K_j^\prime(x)] = -1  \; . \ee

Another sum that we need is
     \be  Q_2  \equiv \lim_{J\rightarrow \infty}  \sum_{j=1/2}^J \left\{(j{+}\frc12)\left[I_j(x)K_j(x)+I_{j+1}(x)K_{j+1}(x)\right]-1\right\}  \;. \ee
It can be obtained by first shifting the indices on the second term and then combining the two terms so that
\bea  Q_2
    &=& \lim_{J \rightarrow \infty} \left\{ \sum_{j=1/2}^J \left[(j{+}\frc12)I_j(x)K_j(x)-1\right] + \sum_{j=3/2}^{J+1} (j{-}\frc12)I_j(x)K_j(x) \right\} \nn \\
   &=& \lim_{J \rightarrow \infty} \left\{ \sum_{j=1/2}^J \left[2jI_j(x)K_j(x)-1\right] + (J{+}\frc12)I_{J+1}(x)K_{J+1}(x) \right\}  \; .
\eea
The limit of the sum is given by Eq.~(\ref{sum1b}), and the limit of the last term can be evaluated using Eqs.~(\ref{smallx}), with the result that
\be\label{sum1d}
   Q_2 =  \sum_{j=1/2}^\infty \left\{(j{+}\frc12)\left[I_j(x)K_j(x)+I_{j+1}(x)K_{j+1}(x)\right]-1\right\} = \frc12 - x \; .
\ee

A slightly more difficult sum is
\be\label{sum1cstep}
   Q_3 \equiv \lim_{J \rightarrow \infty} \sum_{j=1/2}^J \left\{(j{+}\frc12)^2\left[I_j(x)K_j(x)-I_{j+1}(x)K_{j+1}(x)\right]-\frc12\right\} \;.
\ee
As before we begin by shifting indices and combining terms, so that
\bea
\label{sum1cstep2}
      Q_3 &=& \lim_{J\rightarrow \infty} \left\{ \sum_{j=1/2}^J \bigl[(j{+}\frc12)^2I_j(x)K_j(x) - (j{-}\frc12)^2I_j(x)K_j(x) -\frc12\bigr] -(J{+}\frc12)^2 I_{J+1}(x)K_{J+1}(x) \right\} \; . \nn \\ &&
\eea
The limit of the last term must be handled carefully.  With the help of the recursion relations~(\ref{recursion}), the Wronskian~(\ref{Wronsk2}), and the limiting forms~(\ref{smallx}),
this term can be rewritten as
\bea\label{IKlim}
    (J{+}\frc12)^2 I_{J+1}(x)K_{J+1}(x) &=& {x (J{+}\frc12)^2\over 2(J{+}1)} \left[I_J(x)-I_{J+2}(x)\right] K_{J+1}(x)\nonumber \\
    &=& \frc12\left(J + {1 \over 4J{+}4}\right) \left[1 - xI_{J+1}(x) K_J(x) - xI_{J+2}(x)K_{J+1}(x) \right] \nonumber \\
    &=& \frc12 J + O(J^{-1}) = \sum_{j=1/2}^J\frc12 - \frc14 + O(J^{-1}) \; .
\eea
Substituting this into Eq.~(\ref{sum1cstep2}) gives
\be
    Q_3  = \lim_{J\rightarrow \infty} \sum_{j=1/2}^J \left[2jI_j(x)K_j(x) - 1 \right] +\frc14  \; .
\ee
Then using the identity~(\ref{sum1b}) one finds that
\be\label{sum1c}
  Q_3 = \sum_{j=1/2}^\infty \left\{(j{+}\frc12)^2\left[I_j(x)K_j(x)-I_{j+1}(x)K_{j+1}(x)\right]-\frc12\right\} = \frc14 - x \; .
\ee
Multiplying this sum by $x$ and using the recursion relations~(\ref{recursion}) to rewrite $I_j$ in terms of $I_{j+1}$, and $K_{j+1}$ in terms of $K_j$ yields
\be
    \sum_{j=1/2}^\infty \left\{(j{+}\frc12)^2\left[xI_{j+1}^\prime(x)K_j(x)+ xI_{j+1}(x)K_j^\prime(x) + I_{j+1}(x)K_j(x)\right]-\frc12 x\right\} = \frc14x - x^2 \; .
\ee
The left side is a total derivative, so we can integrate and multiply by a factor of $2$ to obtain the identity
\be\label{sum2a}
    \sum_{j=1/2}^\infty \left\{2(j{+}\frc12)^2 xI_{j+1}(x)K_j(x)-\frc12 x^2\right\} = \frc14x^2 - \frc23x^3 \; .
\ee

Another quantity we need is
\be
     Q_4 \equiv \sum_{j=1/2}^\infty \left[j^3I_j(x)K_j(x) - j^2 + \frc12 x^2 \right] \; .
\ee
Its value can be derived from Eq.~(\ref{sum2a}) using Eq.~(\ref{Wronsk2}) in a slightly different way:
\be
    \sum_{j=1/2}^\infty \left\{(j^2+2j+\frc34) xI_{j+1}(x)K_j(x) +  (j^2 - \frc14) \left[1-xI_j(x)K_{j+1}(x) \right]-\frc12 x^2\right\} = \frc14x^2 - \frc23x^3 \; .
\ee
Writing the infinite sum as the limit of a finite sum and then shifting the index so that $j \rightarrow j-1$ on the first term yields
\bea
     \lim_{J \rightarrow \infty}\Biggl[ \sum_{j=1/2}^J \left\{ (j^2-\frc14) \left[xI_j(x)K_{j-1}(x) + 1 - xI_j(x)K_{j+1}(x) \right]-\frc12 x^2\right\} && \nn \\
     +(J^2 + 2J + \frc34) x I_{J+1}(x) K_J(x) \Biggr] &=& \frc14x^2 - \frc23x^3 \; .
\eea
Next use the recursion relations~(\ref{recursion}) to eliminate $K_{j \pm 1}$ and use the limiting form~(\ref{smallx}) on the last term to obtain the identity
\be
     \sum_{j=1/2}^\infty \left\{ (j^2-\frc14) \left[ 1 - 2jI_j(x)K_j(x) \right]-\frc12 x^2\right\} = - \frc23x^3 \; .
\ee
Using the result~\eqref{sum1b} gives the identity
\be\label{sum3a}
     Q_4 = \sum_{j=1/2}^\infty  \left[ 2j^3I_j(x)K_j(x) - j^2 + \frc12x^2 \right] = \frc23x^3 - \frc14 x \;.
\ee
Another identity is obtained from the derivative of this equation:
\be\label{sum3b}
     \sum_{j=1/2}^\infty  \left[ 2j^3I^\prime_j(x)K_j(x) + 2j^3I_j(x)K^\prime_j(x) + x \right] = 2x^2 - \frc14  \; .
\ee

It is also necessary to evaluate integrals of products of Bessel functions and in some cases both integrals and sums of products of Bessel functions.
One such integral is
     \be Q_5 \equiv (2j{+}1) \int_x^\infty \left[K_j^2(y)+K_{j+1}^2(y)\right] dy \;. \ee
Using the recursion relations~(\ref{recursion}) to rewrite $2j K_j(y)$ in terms of $K'_j(y)$ and $K_{j+1}(y)$ and $(2j+2)K_{j+1}(y)$ in terms of $K'_{j+1}(y)$ and $K_j(y)$ one finds
\be
     Q_5  = \int_x^\infty \left[ K_j^2(y) + 2yK_j(y)K_j^\prime(y) - K_{j+1}^2(y) - 2yK_{j+1}(y)K_{j+1}^\prime(y) \right] \; .
\ee
The right side is a total derivative, so
\be\label{Kint}
     Q_5 =  (2j{+}1)\int_x^\infty \left[K_j^2(y)+K_{j+1}^2(y)\right] dy = xK_{j+1}^2(x)-xK_j^2(x) \; .
\ee
The identity
\be\label{Iint}
    (2j{+}1)\int_0^x \left[I_j^2(y)+I_{j+1}^2(y)\right] dy = xI_j^2(x)-xI_{j+1}^2(x)-{2 \over \pi}\delta_{j,-1/2} \; ,
\ee
which is valid for $j \ge -\frc12$, can be derived in a similar way.  The last term comes from the lower limit in the special case $j = -\frc12$.

Next consider
     \be Q_6 \equiv  \sum_{j=1/2}^\infty 4j\left[I_j^2(x)\int_x^\infty K_j^2(y)dy + K_j^2(x)\int_0^x I_j^2(y)dy\right]  \; .\ee
To evaluate this quantity first take the product of Eqs. (\ref{Iint}) and (\ref{Kint}) and sum over $j$ to obtain the identity
\bea
    \lim_{J \rightarrow \infty} \sum_{j=1/2}^J \biggl\{x\left[I_j^2(x)-I_{j+1}^2(x)\right]\int_x^\infty \left[K_j^2(y)+K_{j+1}^2(y)\right]dy \quad && \nn\\
    + x\left[K_j^2(x)-K_{j+1}^2(x)\right]\int_0^x \left[I_j^2(y)+I_{j+1}^2(y)\right]dy \biggr\} &=& 0  \; .
\eea
Then make the index shift $j \rightarrow j-1$ on each of the second terms in the integrals.  This adds additional terms at the lower and upper limits of the sum with the result that
\bea\label{Q5a}
    &&\lim_{J \rightarrow \infty} \sum_{j=1/2}^J \left\{x\left[I_{j-1}^2(x)-I_{j+1}^2(x)\right]\int_x^\infty K_j^2(y) dy + x\left[K_{j-1}^2(x)-K_{j+1}^2(x)\right]\int_0^x I_j^2(y) dy \right\} \nn \\
     && {} = x\left[I_{-1/2}^2(x)-I_{1/2}^2(x)\right] \int_x^\infty K_{1/2}^2(y) dy + x\left[K_{-1/2}^2(x)-K_{1/2}^2(x)\right] \int_0^x I_{1/2}^2(y) dy \nn \\
     && \quad {} + x\lim_{J \rightarrow \infty} \left\{ \left[I_{J+1}^2(x)-I_J^2(x)\right] \int_x^\infty K_{J+1}^2(y) dy + \left[K_{J+1}^2(x) - K_J^2(x)\right] \int_0^x I_{J+1}^2(y)dy \right\} \; . \qquad
\eea
Because we are taking the limit $J \rightarrow \infty$, we can use the approximate Eqs.~\eqref{smallx} to evaluate the $I_J^2$ integral:
\be\label{Isquareint}
     \int_0^x I_J^2(y) dy = \int_0^x {dy \over \Gamma^2(J{+}1)} \left( y \over 2 \right)^{2J} \left[1 + O\left(y^2 \over J \right) \right]  = {2 \over (2J{+}1) \Gamma^2(J{+}1)}  \left( x \over 2 \right)^{2J+1} \left[ 1 + O \left( x^2 \over J \right) \right] \; .
\ee
Eq.~\eqref{Isquareint} is valid not only for large $J$, but whenever $x \ll \sqrt{J}$.

A similar identity can be derived for the $K_J^2$ integral, but because the integral extends to infinity, we must only use the approximate Eqs.~\eqref{smallx} up to some limit $\Lambda$, where $\Lambda \ll \sqrt{J}$.  Hence we first rewrite the integral as
\bea\label{Ksquareint1}
    \int_x^\infty K_J^2(y) dy &=& {\Gamma^2(J) \over 4} \int_x^\Lambda dy \left(2 \over y\right)^{2J} \left[1 + O \left( y^2 \over J\right) \right] + \int_\Lambda^\infty K_J^2(y) dy \nn \\
    &=& {\Gamma^2(J) \over 4J{-}2}\left( 2 \over x \right)^{2J-1} \left\{ 1 + O \left( x^2 \over J \right) + O \left[\left(x\over \Lambda\right)^{2J-1} \right] \right\} + \int_\Lambda^\infty K_J^2(y) dy \; .
\eea
Then the last term can be bounded using Eq.~\eqref{Kint} with $j = J-1$, so that
\bea
     \int_\Lambda^\infty K_J^2(y) dy &=& {\Lambda \over 2J{-}1}  \left[ K_J^2(\Lambda) - K_{j-1}^2(\Lambda) \right] - \int_\Lambda^\infty K_{J-1}^2(y) dy \nn \\
     &<& {\Lambda \over 2J{-}1} K_J^2(\Lambda) \nn \\
     &=& {\Gamma^2(J) \over 4J{-}2} \left( 2 \over x \right)^{2J-1}  \left(x\over \Lambda\right)^{2J-1} \left[ 1 + O\left(\frac{\Lambda^2}{J}\right) \right]   \; .
\eea
If we choose $\Lambda$ noticeably larger than $x$, but still small compared to $\sqrt{J}$, then we can neglect both contributions that are suppressed by $O[(x / \Lambda)^{2J-1}]$, so we simplify Eq.~\eqref{Ksquareint1} to
\be\label{Ksquareint}
    \int_x^\infty K_J^2(y) dy = {\Gamma^2(J) \over (4J{-}2)(x/2)^{2J-1}}\left[ 1 + O \left( x^2 \over J \right) \right] \; .
\ee
Again, Eq.~\eqref{Ksquareint} is valid not only for large $J$, but for any $x \ll \sqrt{J}$ if $J \ge \frc32$.  Combining Eqs.~\eqref{Isquareint} and \eqref{Ksquareint} with Eqs.~\eqref{smallx} gives
\be
    \lim_{J \rightarrow \infty} \left\{ \left[I_{J+1}^2(x)-I_J^2(x)\right] \int_x^\infty K_{J+1}^2(y) dy \right\} = \lim_{J \rightarrow \infty} \left\{ \left[K_{J+1}^2(x) - K_J^2(x)\right] \int_0^x I_{J+1}^2(y)dy \right\} = 0 \; .
\ee
Substituting these two limits into Eq.~\eqref{Q5a} yields
\bea
    &&\lim_{J \rightarrow \infty} \sum_{j=1/2}^J \left\{x\left[I_{j-1}^2(x)-I_{j+1}^2(x)\right]\int_x^\infty K_j^2(y) dy + x\left[K_{j-1}^2(x)-K_{j+1}^2(x)\right]\int_0^x I_j^2(y) dy \right\} \nn \\
     &&\quad {} = x\left[I_{-1/2}^2(x)-I_{1/2}^2(x)\right] \int_x^\infty K_{1/2}^2(y) dy + x\left[K_{-1/2}^2(x)-K_{1/2}^2(x)\right] \int_0^x I_{1/2}^2(y) dy \; .
\eea
Using the recursion relations Eqs. (\ref{recursion}) to rewrite the factors on the left, and using Eqs. (\ref{besselsmall}) to rewrite the factors on the right gives
\be\label{stepA3}
    \sum_{j=1/2}^\infty 4j\left[I_j(x)I'_j(x)\int_x^\infty K_j^2(y) dy + K_j(x)K'_j(x)\int_0^x I_j^2(y) dy \right] =  \int_x^\infty {dy \over y}e^{-2y} \; .
\ee
It is not hard to see that the left side is a total derivative:
\be
    {d \over dx}\sum_{j=1/2}^\infty 2j\left[I_j^2(x)\int_x^\infty K_j^2(y)dy + K_j^2(x)\int_0^x I_j^2(y)dy\right]  = \int_x^\infty {dy \over y}e^{-2y} \; .
\ee
Next multiply by two and integrate from 0 to $x$.  On the left side, it is not hard to show using Eqs.~\eqref{smallx}, \eqref{Isquareint}, and \eqref{Ksquareint} that in the limit $x \rightarrow 0$ all the terms vanish for $j \ge \frc32$.  The $j = \frc12$ term can also be shown to vanish in this limit, so we find
\bea\label{identA}
    Q_6 = \sum_{j=1/2}^\infty 4j\left[I_j^2(x)\int_x^\infty K_j^2(y)dy + K_j^2(x)\int_0^x I_j^2(y)dy\right] = 2x \int_x^\infty {dy \over y} e^{-2y} + 1 - e^{-2x} \; .
\eea

More complicated expressions can be evaluated using Eq.~(\ref{identA}).  For example,
\bea
    Q_7 \equiv \lim_{J\rightarrow \infty} \sum_{j=1/2}^J \biggl\{ 1 + (j{+}\frc12) x \left[I_j^2(x) - I_{j+1}^2(x)\right] \int_x^\infty \left[ K_j^2(y) - K_{j+1}^2(y)\right] dy \qquad &&\nn \\
     {} + (j{+}\frc12)x\left[K_j^2(x) - K_{j+1}^2(x)\right] \int_0^x \left[I_j^2(y)-I_{j+1}^2(y)\right] dy\biggr\}  \; . &&
\eea
 First shift the sums such that the factor in front of the integral is always $I_j^2(x)$ or $K_j^2(x)$.  As usual, this will introduce terms coming from the shifted sum.  The  $j=\frc12$ terms are
 proportional to $j-\frc12$ and thus vanish.  Eqs.~(\ref{smallx}), \eqref{Isquareint}, and \eqref{Ksquareint} can then be used to simplify the terms at large $J$:
\bea
    Q_7 &=& \lim_{J \rightarrow \infty} \Biggl[ \sum_{j=1/2}^J \biggl\{ 1 +xI_j^2(x) \int_x^\infty \left[2jK_j^2(y)-(j{+}\frc12)K_{j+1}^2(y) - (j{-}\frc12)K_{j-1}^2(y)\right] dy \nn \\
    &&\qquad {}+  x K_j^2(x) \int_0^x \left[2jI_j^2(y)-(j{+}\frc12)I_{j+1}^2(y) - (j{-}\frc12) I_{j-1}^2(y) \right] dy \biggr\} \nn \\
    &&\qquad {} -  (J{+}\frc12)x\left\{I_{J+1}^2(x)  \int_x^\infty \left[ K_J^2(y) - K_{J+1}^2(y)\right] dy + K_{J+1}^2(x)\int_0^x \left[I_J^2(y) - I_{J+1}^2(y) \right] dy \right\} \Biggr] \nn \\
    &=& \sum_{j=1/2}^\infty \biggl\{ 1 +xI_j^2(x) \int_x^\infty \left[2jK_j^2(y)-(j{+}\frc12)K_{j+1}^2(y) - (j{-}\frc12)K_{j-1}^2(y)\right] dy \nn \\
    &&\qquad {} + x K_j^2(x) \int_0^x \left[2jI_j^2(y)-(j{+}\frc12)I_{j+1}^2(y) - (j{-}\frc12) I_{j-1}^2(y) \right] dy \biggr\} -\frc12 \; .
\eea
 Using Eqs. (\ref{Kint}) and (\ref{Iint}) the expression can be rewritten as
\bea
    Q_7 &=& \sum_{j=1/2}^\infty \biggl\{ 1 +4jxI_j^2(x) \int_x^\infty K_j^2(y) dy +4j x K_j^2(x) \int_0^x I_j^2(y)dy  + \frc12x^2\bigl[ I_j^2(x)K_{j-1}^2(x) \nn \\
    &&{} - I_{j-1}^2(x)K_j^2(x) + I_{j+1}^2(x)K_j^2(x) - I_j^2(x)K_{j+1}^2(x) \bigr] \biggr\} + {x\over \pi} K_{1/2}^2(x) -\frc12 \; .
\eea
The identity (\ref{identA}) can be used to evaluate the terms with integrals.  The rest of the terms in the sum can be simplified with the help of Eq. (\ref{Wronsk2}).  Then
 using the explicit form for $K_{1/2}$ in Eq. (\ref{besselsmall}) gives
\be
    Q_7 =2x^2 \int_x^\infty {dy \over y} e^{-2y} + x - xe^{-2x} + x\sum_{j=1/2}^\infty \left[ I_j(x)K_{j-1}(x) + I_{j+1}(x)K_j(x) \right] + \frc12e^{-2x} -\frc12 \; .
\ee
This can be simplified using Eq. (\ref{sum1a}) to yield
\bea\label{identB}
    Q_7 &=& \sum_{j=1/2}^\infty \biggl\{ 1 + (j{+}\frc12) x \left[I_j^2(x) - I_{j+1}^2(x)\right] \int_x^\infty \left[ K_j^2(y) - K_{j+1}^2(y)\right] dy  \nn \\
     && \qquad {} + (j{+}\frc12) x\left[K_j^2(x) - K_{j+1}^2(x)\right] \int_0^x \left[I_j^2(y)-I_{j+1}^2(y)\right] dy\biggr\} \nn \\
     &=& 2x^2 \int_x^\infty {dy \over y} e^{-2y} + 2x - \frc12 - xe^{-2x} + \frc12e^{-2x} \; .
\eea

Another expression we will need to evaluate is
\bea
    Q_8 \equiv \sum_{j=1/2}^\infty \biggl[2x(j{+}\frc12)^2\biggl\{I_j(x)I_{j+1}(x) \int_x^\infty\left[K_j^2(y)-K_{j+1}^2(y)\right]dy \qquad\qquad && \nn \\
    {} - K_j(x)K_{j+1}(x) \int_0^x \left[I_j^2(y) - I_{j{+}1}^2(y) \right] dy \biggr\} + x \biggr] \; . &&
\eea
 Begin by taking the derivative of this expression.  Then use the recursion relations Eq.~(\ref{recursion}) to eliminate the derivatives of the Bessel functions.  Next use Eq.~(\ref{Wronsk2}) to obtain
\bea
    {d \over dx} Q_8 &=& \lim_{J\rightarrow \infty} \sum_{j=1/2}^J \biggl[ 2(j{+}\frc12)^2\biggl\{x\left[I_j^2(x)+I_{j+1}^2(x)\right] \int_x^\infty \left[K_j^2(y)-K_{j+1}^2(y)\right] dy \nn \\
   &+&x\left[K_j^2(x)+K_{j+1}^2(x)\right] \int_0^x \left[I_j^2(y)-I_{j+1}^2(y)\right] dy + I_{j+1}(x)K_{j+1}(x) - I_j(x)K_j(x)\biggr\} + 1 \biggr]  \; . \nn \\ &&
\eea
 Eq.~(\ref{sum1c}) can be used to simplify the nonintegrated terms in the sum.  Then shift the sums so the factor in front of the integral is always $I_j^2(x)$ or $K_j^2(x)$.  This shift does not introduce any new terms at small $j$, but it does at large $J$.  The result is
\bea\label{stepC2}
   {d \over dx} Q_8 &=& \lim_{J\rightarrow \infty} \Biggl\{\sum_{j=1/2}^J \biggl[2xI_j^2(x) \int_x^\infty \left\{(j{+}\frc12)^2\left[K_j^2(y)-K_{j+1}^2(y)\right] +(j{-}\frc12)^2 \left[K_{j-1}^2(y)-K_j^2(y)\right] \right\}dy  \nn \\
    &&\qquad \qquad \qquad {}+ 2xK_j^2(x) \int_0^x \left\{ (j{+}\frc12)^2 \left[I_j^2(y)-I_{j+1}^2(y)\right] + (j{-}\frc12)^2 \left[I_{j-1}^2(y)-I_j^2(y)\right] \right\} dy  \biggr] \nn \\
    &+& 2(J{+}\frc12)^2x\left\{I_{J+1}^2(x)\int_x^\infty \left[ K_J^2(y)-K_{J+1}^2(y)\right] dy + K_{J+1}^2(x) \int_0^x \left[I_J^2(y) - I_{J+1}^2(y)\right] dy \right\} \Biggr\} \; \nn \\
    &&{} + 2x - \frc12 \; .
\eea
The large $J$ terms, with the help of Eqs.~\eqref{smallx}, \eqref{Isquareint} and \eqref{Ksquareint}, can all be seen to vanish except for one, which can be computed with the additional help of Eq.~\eqref{Iint}:
\bea
    &&2(J{+}\frc12)^2 x K_{J+1}^2(x) \int_0^x I_J^2(y) dy  =  x(J{+}\frc12) K_{J+1}^2(x)\left[xI_J^2(x) - xI_{J+1}^2(x) - (2J{+}1) \int_0^x I_{J+1}^2(y) dy \right] \nn \\
    && \qquad = (J{+}\frc12)\left\{\left[1-xI_{J+1}(x)K_J(x)\right]^2 - x^2K_{J+1}^2(x)I_{J+1}^2(x)- (2J{+}1) xK_{J+1}^2(x) \int_0^x I_{J+1}^2(y) dy  \right\} \nn \\
    && \qquad = J {+} \frc12 +O(J^{-1}) \; .
\eea
Here Eq.~(\ref{Wronsk2}) was used to help approximate the leading term.  Substituting this in Eq.~(\ref{stepC2}) gives
\bea
     {d \over dx}Q_8 &=&\lim_{J \rightarrow \infty} \Biggl\{ \sum_{j=1/2}^J \biggl[2xI_j^2(x)\int_x^\infty \left\{(j{+}\frc12)^2\left[K_j^2(y)-K_{j+1}^2(y)\right] + (j{-}\frc12)^2\left[K_{j-1}^2(y)-K_j^2(y)\right] \right\} dy  \nn \\
     && {} + 2xK_j^2(x) \int_0^x  \left\{ (j{+}\frc12)^2 \left[I_j^2(y) - I_{j+1}^2(y)\right] + (j{-}\frc12)^2 \left[I_{j-1}^2(x) - I_j^2(x) \right] \right\} dy \biggr] + (J{+}\frc12) \Biggr\} \nn \\
      && {} + 2x-\frc12 \; .
\eea
Next use Eqs.~(\ref{Kint}) and (\ref{Iint}) in such a way that only the terms $I_j^2(y)$ or $K_j^2(y)$ appear inside the integrals.  The result is
\bea
     {d \over dx}Q_8 &=& 2x - \frc12 + \lim_{J \rightarrow \infty} \Biggl[  (J{+}\frc12) + \sum_{j=1/2}^J \biggl\{8xjI_j^2(x)\int_x^\infty K_j^2(y)dy + 8xjK_j^2(x) \int_0^x I_j^2(y) dy \nn \\
     {} &+& x^2 (j{+}\frc12)\left[I_{j+1}^2(x)K_j^2(x)-I_j^2(x)K_{j+1}^2(x)\right] \nonumber + x^2 (j{-}\frc12) \left[I_{j-1}^2(x)K_j^2(x)-I_j^2(x)K_{j-1}^2(x) \right] \biggr\} \Biggr] \, . \nn \\ &&
\eea
The remaining integrals are given by Eq.~(\ref{identA}).  The two terms in the final line in the sum are equal and opposite save for a shift of index. Thus
\be\label{stepC3}
     {d \over dx} Q_8  = 4x^2\int_x^\infty {dy \over y} e^{-2y} + 4x - 2xe^{-2x} - \frc12 + \lim_{J\rightarrow \infty} (J{+}\frc12)\left\{1+x^2  \left[I_{J+1}^2(x)K_J^2(x) - I_J^2(x)K_{J+1}^2(x) \right]\right\}  \; .
\ee
With the help of Eq.~(\ref{Wronsk2}) and then Eq.~(\ref{smallx}), the remaining limit is
\be
     \lim_{J\rightarrow \infty} (J{+}\frc12) \left\{1+x^2 \left[I_{J+1}^2(x)K_J^2(x) - I_J^2(x)K_{J+1}^2(x) \right]\right\} = \lim_{J\rightarrow \infty} (2J{+}1)x I_{J+1}(x) K_J(x) = 0 \; .
\ee
After integrating both sides of Eq.~(\ref{stepC3}) from 0 to $x$, one can use the limiting forms Eqs.~\eqref{smallx}, \eqref{Isquareint}, and \eqref{Ksquareint} to show that the left side vanishes at $x=0$.  The result is:
\bea\label{identC}
     Q_8 &=& \sum_{j=1/2}^\infty \biggl[2x(j{+}\frc12)^2\biggl\{I_j(x)I_{j+1}(x) \int_x^\infty\left[K_j^2(y) - K_{j+1}^2(y)\right]dy  \nn \\
     &&\qquad {} - K_j(x)K_{j+1}(x) \int_0^x \left[I_j^2(y)  -  I_{j+1}^2(y) \right] dy \biggr\} + x \biggr]  \nn \\
     &=&  \frc43 x^3 \int_x^\infty {dy\over y}e^{-2y} -\frc23x^2e^{-2x}+\frc13xe^{-2x}+\frc16e^{-2x} +2x^2-\frc12x-\frc16 \; .
\eea

The final quantity which we need to evaluate is
     \be\label{Q8def} Q_9 \equiv \sum_{j=1/2}^\infty  \left[ 4j^3 I_j^2(x) \int_x^\infty K_j^2(y) dy + 4j^3 K_j^2(x) \int_0^x I_j^2(y) dy  - x \right] \; . \ee
To do so first rewrite Eq.~(\ref{identC}) as the limit of a finite sum, and then shift $j \rightarrow j-1$ on the terms with integrals that contain $K_{j+1}^2$ or $I_{j+1}^2$ to yield
\bea\label{stepD1}
     &&\lim_{J \rightarrow \infty} \Biggl[ \sum_{j=1/2}^J \biggl\{ 2x I_j(x) \left[(j{+}\frc12)^2 I_{j{+}1}(x) - (j{-}\frc12)^2 I_{j{-}1}(x) \right] \int_x^\infty K_j^2(y) dy  \nn \\
     &&\qquad\qquad {}+ 2x K_j(x)\left[ (j{-}\frc12)^2 K_{j-1}(x) - (j{+}\frc12)^2 K_{j{+}1}(x)\right]\int_0^x I_j^2(y) dy + x \biggr\} \nn \\
     && \qquad\qquad {}- 2x (J{+}\frc12)^2 I_J(x) I_{J{+}1}(x) \int_x^\infty K_{J{+}1}^2(y) dy + 2x (J{+}\frc12)^2 K_J(x) K_{J+1}(x) \int_0^x I_{J+1}^2(y) dy  \Biggr] \nn \\
     && \hbox{\hskip1in} {} = \frc43 x^3 \int_x^\infty {dy\over y}e^{-2y} -\frc23x^2e^{-2x}+\frc13xe^{-2x}+\frc16e^{-2x} +2x^2-\frc12x-\frc16 \; .
\eea
The large $J$ terms can be evaluated with the help of Eqs.~(\ref{smallx}), \eqref{Isquareint}, and \eqref{Ksquareint}, and the recursion relations \eqref{recursion} can be used to rewrite this as
\bea
     &&\sum_{j=1/2}^\infty \biggl\{  \left[ 4jx I_j(x) I_j^\prime(x) - (4j^3+j) I_j^2(x) \right] \int_x^\infty K_j^2(y) dy \nn \\
     && \qquad {} +\left[ 4jx K_j(x) K_j^\prime(x) - (4j^3+j) K_j^2(x) \right] \int_0^x I_j^2(y) dy + x \biggr\} \nn \\
     && \qquad \qquad {} = \frc43 x^3 \int_x^\infty {dy\over y}e^{-2y} -\frc23x^2e^{-2x}+\frc13xe^{-2x}+\frc16e^{-2x} +2x^2-\frc16 \; .
\eea
The terms with derivatives can be evaluated using Eq.~(\ref{stepA3}) and the remaining terms proportional to $j$ can be evaluated using Eq.~(\ref{identA}) with the result that
\bea\label{identD}
    Q_9  &=&\sum_{j=1/2}^\infty  \left[ 4j^3 I_j^2(x) \int_x^\infty K_j^2(y) dy + 4j^3 K_j^2(x) \int_0^x I_j^2(y) dy  - x \right] \nn \\
      &=& \left( \frc12x - \frc43 x^3\right) \int_x^\infty {dy\over y}e^{-2y} + \frc23x^2e^{-2x} - \frc13xe^{-2x} + \frc1{12}e^{-2x} - 2x^2-\frc1{12} \; .
\eea

\end{document}